\documentclass[twocolumn,prb,showpacs,preprintnumbers,superscriptaddress,amsmath,amssymb,citeautoscript]{revtex4-2}
\usepackage{graphicx}
\usepackage{epstopdf}
\usepackage{dcolumn}
\usepackage{color}
\usepackage{amsmath,amssymb}
\usepackage[english]{babel}
\usepackage{bm}

\usepackage[hyperindex]{hyperref}
\hypersetup{colorlinks,citecolor=blue, filecolor=blue,linkcolor=blue,urlcolor=blue}

\usepackage{multirow}
\usepackage{rotating}
\usepackage{array}

\newcommand{\bpm}{\begin{pmatrix}}
\newcommand{\epm}{\end{pmatrix}}
\newcommand{\bs}{\boldsymbol}
\newcommand{\be}{\begin{equation}}
\newcommand{\ee}{\end{equation}}
\newcommand{\beq}{\begin{eqnarray}}
\newcommand{\eeq}{\end{eqnarray}}

\DeclareMathOperator{\im}{Im}

\DeclareMathOperator{\tr}{tr}

\begin{document}

\title{Quasiparticle interference patterns in bilayer graphene with trigonal warping}

\author{Vardan Kaladzhyan}
\email{vardan.kaladzhyan@phystech.edu}
\affiliation{Department of Physics, University of Basel, Klingelbergstrasse 82, CH-4056 Basel, Switzerland}
\author{Fr\'ed\'eric Joucken}
\affiliation{Department of Physics, University of California, Santa Cruz, California, USA}
\author{Zhehao Ge}
\affiliation{Department of Physics, University of California, Santa Cruz, California, USA}
\author{Eberth A. Quezada-Lopez}
\affiliation{Department of Physics, University of California, Santa Cruz, California, USA}
\author{Takashi Taniguchi}
\affiliation{International Center for Materials Nanoarchitectonics, National Institute for Materials Science, 1-1 Namiki, Tsukuba 305-0044, Japan}
\author{Kenji Watanabe}
\affiliation{Research Center for Functional Materials, National Institute for Materials Science, 1-1 Namiki, Tsukuba 305-0044, Japan}
\author{Jairo Velasco Jr}
\affiliation{Department of Physics, University of California, Santa Cruz, California, USA}
\author{Cristina Bena}
\affiliation{Institut de Physique Th\'eorique, Universit\'e Paris Saclay, CEA
CNRS, Orme des Merisiers, 91190 Gif-sur-Yvette Cedex, France}

\date{\today}

\begin{abstract}
We calculate the form of quasiparticle interference patterns in bilayer graphene within a low-energy description, taking into account perturbatively the trigonal warping terms. We introduce four different types of impurities localized on the A and B sublattices of the first and the second layer, and we obtain closed-form analytical expressions both in real and Fourier spaces for the oscillatory corrections to the local density of states generated by the impurities. Finally, we compare our findings with recent experimental and semi-analytical T-matrix results from arXiv:2104.10620 and we show that there is a very good agreement between our findings and the previous results, as well as with the experimental data. 
\end{abstract}

\maketitle

\section{Introduction}

In realistic materials impurities and defects are ubiquitous players that often hinder the interpretation of experimental results. However, certain manifestations of their presence can be  useful to reveal some interesting properties and to access the parameters of the underlying systems. For instance, Friedel oscillations in the local density of states \cite{Friedel1958}, often referred to as ``quasiparticle interference patterns", allow to extract the Fermi momentum of the electrons, as well as to infer about the dimension of the system under consideration. These oscillations arise as a result of impurity scattering processes and represent interferences between incoming and outgoing wave packets.

Quasiparticle interference patterns have been extensively studied, both theoretically and experimentally, in graphene, be it single layer or multi-layer \cite{Cheianov2006,Mariani2007,Kivelson2003,Bena2005,Bena2008,Bena2009,Wehling2007,Peres2006,Peres2007,Vozmediano2005,Ando2006, Pogorelov2006,Skrypnyk2006,Skrypnyk2007,Katsnelson2007,Dutreix2016}. For instance, it was shown that in a single-layer graphene sheet the correction to the local density of states in the presence of impurities decays as $1/r^2$ \cite{Cheianov2006,Mariani2007,Bena2008,Bena2009} versus the expected $1/r$ behavior prevalent in two-dimensional systems \cite{Bena2005,Bena2008}. However, as it was demonstrated in Ref.~[\onlinecite{Bena2008}], the $1/r$ decay is restored in bilayer graphene. 
Experimentally quasiparticle interference patterns are accessible via Fourier-transform scanning tunneling microscopy \cite{Mallet2007,Mallet2016}. 

\begin{figure}
	\centering
	\includegraphics[width=0.49\columnwidth]{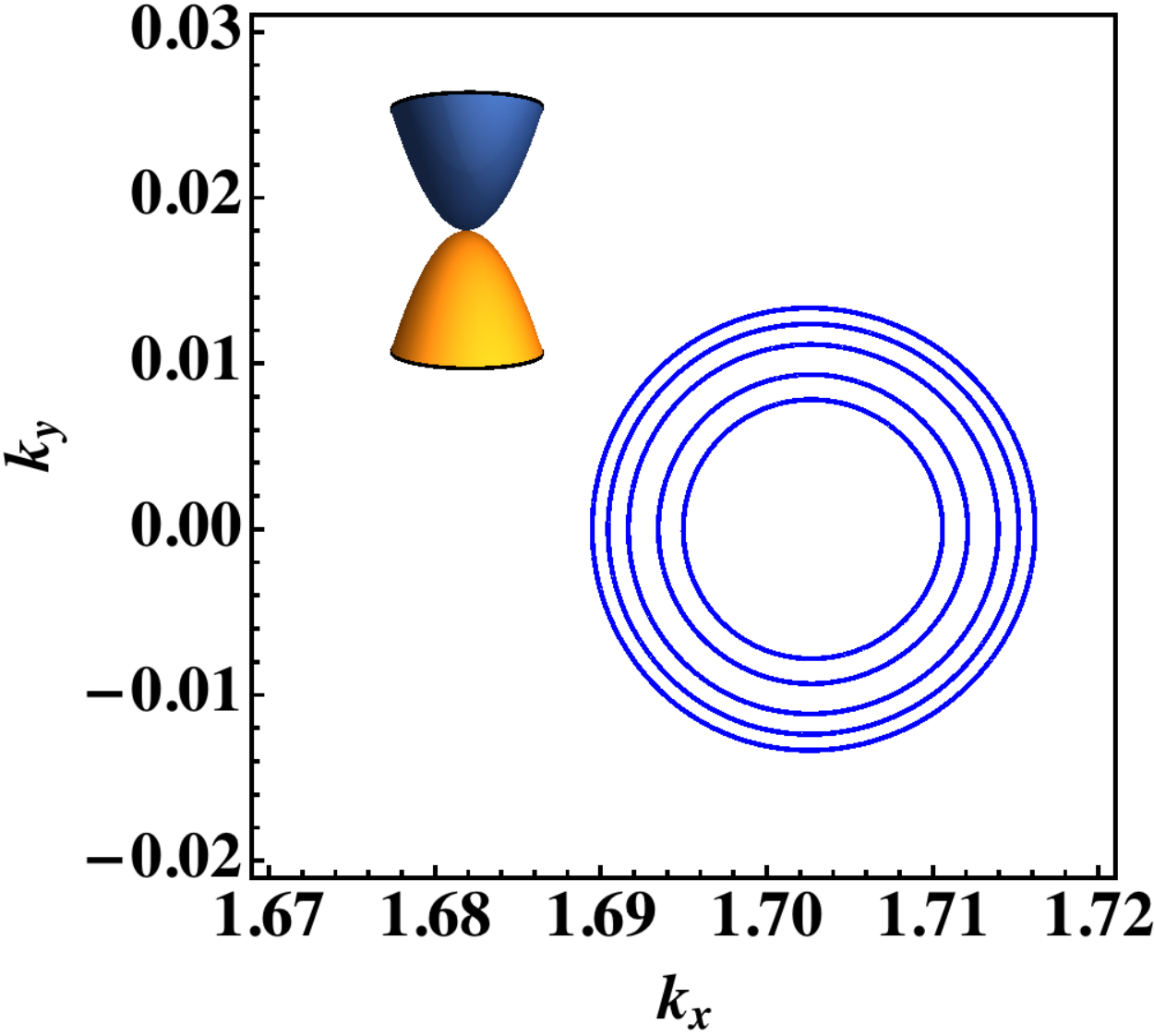}
	\includegraphics[width=0.49\columnwidth]{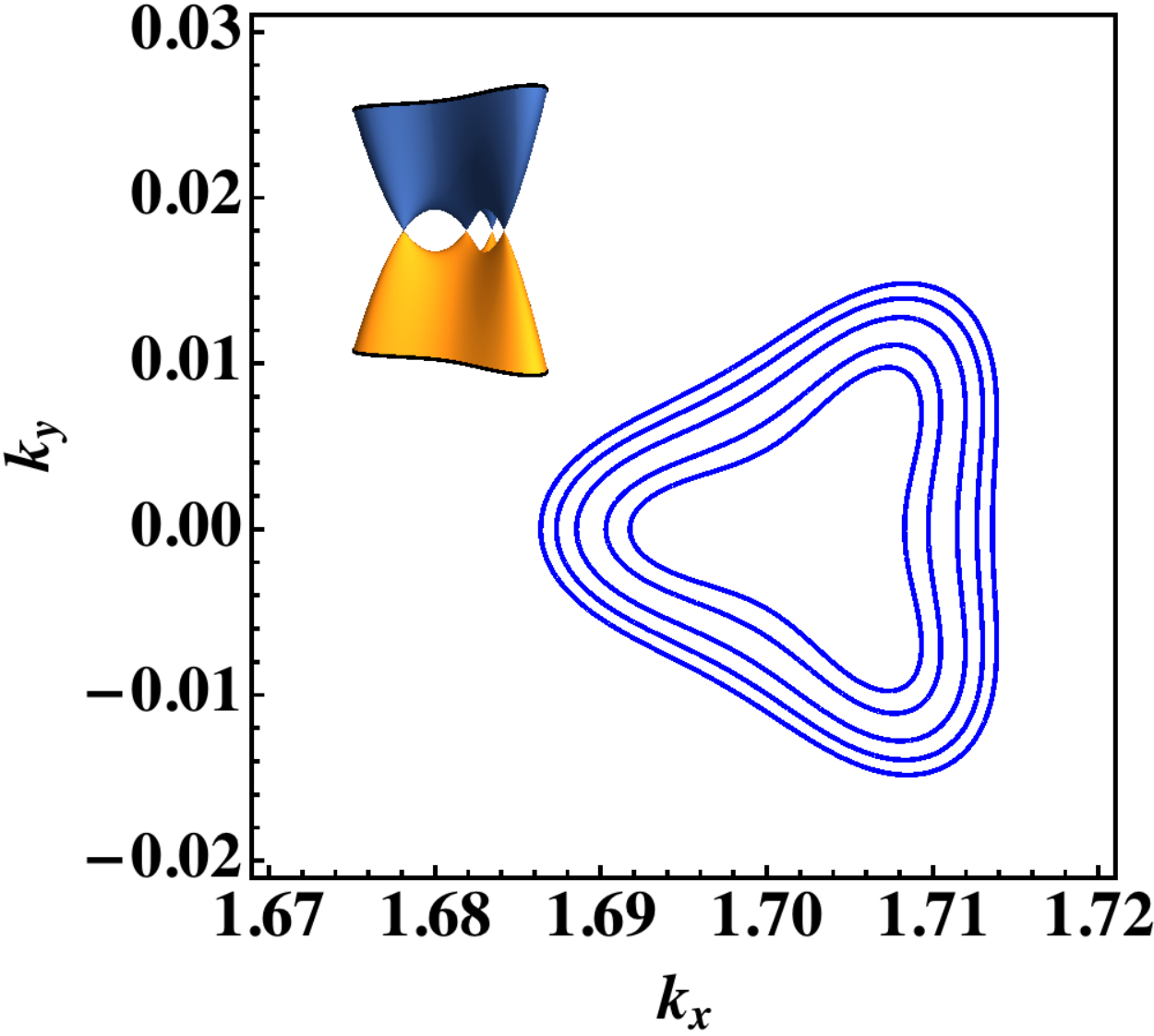}
	\caption{Equal-energy contours for bilayer graphene in the absence and in the presence of trigonal warping (left and right panels, respectively). The corresponding band structures are shown as insets. }
	\label{figTW}
\end{figure}

It was anticipated theoretically \cite{Charlier1991,McCann2006,McCann2013} and confirmed experimentally \cite{Kuzmenko2009,Joucken2020,Zibrov2018,Varlet2014,Orlita2012,Shi2018} that there is a threefold symmetric warping of the bands in bilayer graphene, the so-called ``trigonal warping". Originating from the interlayer coupling between non-dimer orbitals A1 and B2, the latter splits the Dirac points inherent in graphene-like systems into four Dirac points, as it is shown in Fig.~\ref{figTW} for one of the valleys. Furthermore, as it was demonstrated in Refs.~[\onlinecite{Charlier1991,McCann2013,Kechedzhi2007,Zibrov2018,Varlet2014,Orlita2012,Shi2018,Ge2020}], the trigonal warping can have important consequences on the physical properties of graphitic systems. It is worth noting that the effect of the trigonal warping observable in experiments is stronger as one approaches the Dirac points. Finally, note that some confusion regarding the orientation of the trigonal warping has been solved both theoretically \cite{Jung2014} and experimentally \cite{Joucken2020}.

In this paper we calculate analytically the quasiparticle interference patterns in bilayer graphene taking into account the trigonal warping terms. Using a first-order perturbative expansion in trigonal warping, and the $T$-matrix formalism \cite{Byers1993,Salkola1996,Ziegler1996,Mahan2000,Balatsky2006,Bena2016}, we find closed analytical expressions for the correction to the local density of states introduced by four distinct types of impurities localized at different sublattices and layers of the bilayer graphene. Our most interesting observations are that at the energies close to the Dirac points the real-space oscillations reflect the symmetry of the trigonal warping, and most saliently, one can extract in a closed analytical form the values of momenta at which the oscillations occur. By analyzing our findings both in the real and Fourier space, we provide a comparison with recent experiments and semi-analytical calculations \cite{Joucken2020,Joucken2021}, and we show that our results are in good agreement with the experimental observations.

We proceed as follows: in Sec.~II we derive the low-energy description of bilayer graphene, in Sec.~III we calculate perturbatively the bare retarded Green's function of the system in the real and Fourier space. In Sec.~IV we apply the $T$-matrix formalism to compute the  quasiparticle interference patterns and compare these findings to the experimental results, leaving the conclusions to Sec.~V.  

\section{Low-energy description}
We start by writing down the simplest lattice model for bilayer graphene \cite{McCann2013,Joucken2020}. In the basis $\left\{ \psi_{\bs{k}}^{A1},\, \psi_{\bs{k}}^{B1},\, \psi_{\bs{k}}^{A2},\, \psi_{\bs{k}}^{B2} \right\}$, where $A, B$ and $1, 2$ refer to sublattices and layers, respectively, we have:
\begin{align}
\label{H_lattice}
H(\bs{k}) \equiv 
	\bpm
		0 & -\gamma_0 f(\bs{k}) & 0 & -\gamma_3 f^*(\bs{k}) \\
		-\gamma_0 f^*(\bs{k}) & 0 & \gamma_1 & 0 \\
		0 & \gamma_1 & 0 & -\gamma_0 f(\bs{k}) \\
		-\gamma_3 f(\bs{k}) & 0 & -\gamma_0 f^*(\bs{k}) & 0
	\epm,
\end{align}
where we defined $f(\bs{k}) \equiv e^{ik_ya/\sqrt{3}} + 2 e^{-ik_ya/2\sqrt{3}} \cos \frac{k_x a}{2}$, $a = 2.46$ \AA  \, is the lattice constant, the intra- and interlayer hopping constants are denoted by $\gamma_0 = 3.3\,$eV and $\gamma_1 = 0.42\,$eV, respectively, and the trigonal warping parameter $\gamma_3 = - 0.3\,$eV. The sign of the latter stands for the warping orientation. 

To derive a low-energy theory we expand the Hamiltonian in Eq.~(\ref{H_lattice}) around $K_s \equiv \left( s \frac{4\pi}{3a}, 0\right)$ points, where $s = \pm$ is the valley index, and we get:
\begin{align}
f(\bs{k}) \approx -\frac{\sqrt{3}}{2}(s k_x - i k_y)a
\end{align}
To simplify further the calculations, we divide the Hamiltonian above by $\gamma_0 \sqrt{3}/2$, thereby  rendering it dimensionless, and we introduce $\bs{q} \equiv \bs{k} a$:
\begin{align}
\label{H_low_energy}
\mathcal{H}(\bs{q}) = 
	\bpm
		0 & s q e^{-is\phi_q} & 0 & \gamma_{03} s q e^{+is\phi_q} \\
		s q e^{+is\phi_q} & 0 & \gamma_{01} & 0 \\
		0 & \gamma_{01} & 0 & s q e^{-is\phi_q} \\
		\gamma_{03} s q e^{-is\phi_q} & 0 & s q e^{+is\phi_q} & 0 
	\epm,
\end{align}
where
\begin{align}
\mathcal{H} \equiv \frac{2H}{\gamma_0 \sqrt{3}},  \; \gamma_{01} = \frac{2\gamma_{1}}{\gamma_0 \sqrt{3}} \approx 0.15, \; \gamma_{03} =  \frac{\gamma_{3}}{\gamma_0} \approx -0.09.
\end{align}
For the sake of simplicity,  above we introduced polar coordinates in momentum space, i.e., we replaced $\bs{q} = (q_x,\,q_y) \to (q,\, \phi_q)$, where $q =\sqrt{q_x^2  + q_y^2} \geqslant 0$, and $\phi_q \in \left[0,\,2\pi\right)$ with $e^{\pm i s \phi_q} = \frac{q_x \pm i s q_y}{q}$. 

The Hamiltonian in Eq.~(\ref{H_low_energy}) provides a low-energy description for bilayer graphene with trigonal warping.

\section{Perturbative calculation of the retarded Green's function}

In what follows we calculate the bare retarded Green's function in momentum space and in real space. The former is easily feasible, however, the latter requires a very complicated Fourier transform. Therefore, below we adopt a different strategy and we resort to a perturbative approach for both the momentum-space and real-space calculations.

\subsection{Momentum space}
We rewrite the Hamiltonian in Eq.~(\ref{H_low_energy}) as a sum of $\mathcal{H}_0(\bs{q})$, which is unperturbed by trigonal warping, and $\mathcal{V}(\bs{q})$, embodying the trigonal warping:
\begin{align}
\nonumber \mathcal{H}(\bs{q}) =&
\underbrace{ 
	\bpm
		0 & s q e^{-is\phi_q} & 0 & 0 \\
		s q e^{+is\phi_q} & 0 & \gamma_{01} & 0 \\
		0 & \gamma_{01} & 0 & s q e^{-is\phi_q} \\
		0 & 0 & s q e^{+is\phi_q} & 0 
	\epm}_{\mathcal{H}_0(\bs{q})} + \\
&\underbrace{ 
	\bpm
		0 & 0 & 0 & \gamma_{03} s q e^{+is\phi_q} \\
		0 & 0 & 0 & 0 \\
		0 & 0 & 0 & 0 \\
		\gamma_{03} s q e^{-is\phi_q} & 0 & 0 & 0 
	\epm}_{\mathcal{V}(\bs{q})}.
\end{align} 
Below we calculate the Matsubara Green's function $G(i\omega,\bs{q}) \equiv \left[i\omega - \mathcal{H}(\bs{q}) \right]^{-1}$, and then perform an analytic continuation replacing $i\omega \to \epsilon + i0^+$ to recover the retarded Green's function $\mathcal{G}(\epsilon,\bs{q})$. 

A first-order perturbation theory in $\gamma_{03}$ yields:
\begin{align}
\nonumber G(i\omega_n, \bs{q}) = \left[i\omega - \mathcal{H}_0(\bs{q}) - \mathcal{V}(\bs{q}) \right]^{-1} = \phantom{aaaaaaaaaaaa}\\
\nonumber = \left[ G^{-1}_0(i\omega, \bs{q})- \mathcal{V}(\bs{q}) \right]^{-1} \approx G_0(i\omega, \bs{q}) + G_1(i\omega, \bs{q}),
\end{align}
where $G_1(i\omega, \bs{q}) \equiv G_0(i\omega, \bs{q}) \mathcal{V}(\bs{q}) G_0(i\omega, \bs{q})$ is the first-order correction in $\gamma_{03}$. Finally, in momentum space we have:
\begin{align}
G_0(i\omega, \bs{q}) =
	\frac{1}{D_q} 
	\bpm
		g_0^{11} & \left(g_0^{21}\right)^\star & \left(g_0^{31}\right)^\star & \left(g_0^{41}\right)^\star \\
		g_0^{21} & g_0^{22} & g_0^{32} & \left(g_0^{31}\right)^\star \\
		g_0^{31} & g_0^{32} & g_0^{22} & \left(g_0^{21}\right)^\star \\
		g_0^{41} & g_0^{31} & g_0^{21} & g_0^{11}
	\epm.
\end{align}
Here $D_q \equiv (q^2-(i\omega)^2+i\omega \gamma_{01} )(q^2-(i\omega)^2-i\omega\gamma_{01})$, while $ g_0^{ij} $ denotes the $(ij)$-th element of the matrix $g_0$, with
\begin{align}
\nonumber &g_0^{11} \equiv i\omega((i\omega)^2 - \gamma_{01}^2) - i\omega q^2,\;\;
g_0^{22} \equiv (i\omega)^3 - (i\omega) q^2, \\
\nonumber &g_0^{21} \equiv (i\omega)^2 \cdot sqe^{is\phi_q} - s q^3 e^{is\phi_q},\;\; 
g_0^{31} \equiv  i\omega \gamma_{01} \cdot sqe^{is\phi_q},\\
&g_0^{32} \equiv (i\omega)^2 \gamma_{01},\;\; 
g_0^{41} \equiv \gamma_{01} \cdot q^2 e^{2is\phi_q}.
\end{align}
We use the $\star$ symbol to denote replacing $\phi_q \to -\phi_q$, e.g., $\left(g_0^{31}\right)^\star = i\omega \gamma_{01} \cdot sqe^{-is\phi_q}$. 

The first-order correction in trigonal warping is given by
\begin{align}
G_1(i\omega, \bs{q}) = 
	\frac{\gamma_{03}}{D_q^2} 
	\bpm
		g_1^{11} & \left(g_1^{21}\right)^\star & \left(g_1^{31}\right)^\star & \left(g_1^{41}\right)^\star \\
		g_1^{21} & g_1^{22} & \left(g_1^{32}\right)^\star & \left(g_1^{31}\right)^\star \\
		g_1^{31} & g_1^{32} & g_1^{22} & \left(g_1^{21}\right)^\star \\
		g_1^{41} & g_1^{31} & g_1^{21} & g_1^{11}
	\epm,
\end{align}
where $g_1^{ij}$ are defined in Appendix \ref{App:FirstOrder}.

\subsection{Bare retarded Green's function in real space}
In this subsection we calculate the Fourier transform of the Matsubara Green's function obtained perurbatively in the previous section. For this we need to calculate the two following integrals:
\begin{align}
\nonumber G(i\omega,\bs{r}) &= \underbrace{\int\limits_0^{+\infty} \frac{qdq}{2\pi} \int\limits_0^{2\pi} \frac{d\phi_q}{2\pi} G_0(i\omega, \bs{q}) e^{i q r \cos(\phi_q - \phi_r)} }_{G_0(i\omega, \bs{r})} + \\
&\underbrace{\int\limits_0^{+\infty} \frac{qdq}{2\pi} \int\limits_0^{2\pi} \frac{d\phi_q}{2\pi} G_1(i\omega, \bs{q}) e^{i q r \cos(\phi_q - \phi_r)} }_{G_1(i\omega, \bs{r})}.
\label{Grealspace} 
\end{align}
Above we introduced polar coordinates in real space, as well as in momentum space, i.e., $\bs{r} = (x,\, y) \to (r,\,\phi_r)$, where $r \geqslant 0$ and $\phi_r \in \left[0,\,2\pi\right)$. For the sake of brevity, we leave the final real-space form of the Green's functions, as well as the lengthy integral calculations to Appendixes \ref{App:IntegralsZerothOrder} and \ref{App:IntegralsFirstOrder}.


\section{Quasiparticle Interference Patterns}\label{Sec:QPI}
In what follows we introduce localized delta-function impurities into the system and we calculate the associated quasiparticle interference patterns via the $T$-matrix formalism. To simplify the derivation we assume that impurities are localised on a specific sublattice and in a specific layer, and we define their amplitudes as
\begin{align}
\nonumber V_{A1} &= U \bpm
		1 & 0 & 0 & 0 \\
		0 & 0 & 0 & 0 \\
		0 & 0 & 0 & 0 \\
		0 & 0 & 0 & 0 
	\epm\negthickspace,\, 
	\;
V_{B1} = U\bpm
		0 & 0 & 0 & 0 \\
		0 & 1 & 0 & 0 \\
		0 & 0 & 0 & 0 \\
		0 & 0 & 0 & 0 
	\epm\negthickspace, 
	\\
V_{A2} &= U\bpm
		0 & 0 & 0 & 0 \\
		0 & 0 & 0 & 0 \\
		0 & 0 & 1 & 0 \\
		0 & 0 & 0 & 0 
	\epm\negthickspace,\,  
	\;
V_{B2} = U\bpm
		0 & 0 & 0 & 0 \\
		0 & 0 & 0 & 0 \\
		0 & 0 & 0 & 0 \\
		0 & 0 & 0 & 1 
	\epm\negthickspace,
\end{align}
where $U$ denotes the magnitude of the impurity potential. Below we proceed in two steps: first, with the help of the real-space form of the Green's function defined in Eqs.~(\ref{Grealspace}), (\ref{G0RS}) and (\ref{G111}-\ref{G121}), we compute the $T$-matrix that accounts for all-order impurity scattering processes. Second, we find the correction to the local density of states in the presence of impurities.

\subsection{$T$-matrix}
To find the $T$-matrix we need to evaluate the following expression \cite{Byers1993,Salkola1996,Ziegler1996,Mahan2000,Balatsky2006,Bena2016}:
\begin{align}
T(i\omega) = \left[\mathbb{I} - V \cdot \lim\limits_{\bs{r} \to \bs{0}} G(i\omega,\bs{r}) \right]^{-1} V,
\end{align}
where $G(i\omega,\bs{r}) \equiv G_0(i\omega,\bs{r}) + G_1(i\omega,\bs{r})$, and $V = V_{A1}$, $V_{B1}$, $V_{A2}$, $V_{B2}$, depending on the chosen impurity type. The first-order correction to the Green's function does not contribute to the $T$-matrix, since $\lim\limits_{\bs{r} \to \bs{0}} G_1(i\omega,\bs{r}) = 0$ due to the angular parts. Thus, the corresponding T-matrices are given by:
\begin{align}
\nonumber T_{A1} &= \bpm
		f(i\omega) & 0 & 0 & 0 \\
		0 & 0 & 0 & 0 \\
		0 & 0 & 0 & 0 \\
		0 & 0 & 0 & 0 
	\epm \negthickspace,\,
T_{B1} = \bpm
		0 & 0 & 0 & 0 \\
		0 & g(i\omega) & 0 & 0 \\
		0 & 0 & 0 & 0 \\
		0 & 0 & 0 & 0 
	\epm\negthickspace, 
	\\
T_{A2} &=  \bpm
		0 & 0 & 0 & 0 \\
		0 & 0 & 0 & 0 \\
		0 & 0 & g(i\omega) & 0 \\
		0 & 0 & 0 & 0 
	\epm\negthickspace,\,
T_{B2} =  \bpm
		0 & 0 & 0 & 0 \\
		0 & 0 & 0 & 0 \\
		0 & 0 & 0 & 0 \\
		0 & 0 & 0 & f(i\omega) 
	\epm\negthickspace, \label{eq:Tmatrices}
\end{align}
where we defined
\begin{align}
\label{eq:f}f(i\omega) &\equiv \frac{U}{1 - U \lim\limits_{\bs{r} \to \bs{0}} \left[ i\omega((i\omega)^2-\gamma_{01}^2) I_{00} - i\omega I_{03} \right] }, \\
\label{eq:g} g(i\omega) &\equiv \frac{U}{1 - U \lim\limits_{\bs{r} \to \bs{0}} \left[ (i\omega)^3 I_{00} - i\omega I_{03} \right] }.
\end{align}
We leave the calculation of the limits in Eqs.~(\ref{eq:f}) and (\ref{eq:g}) to Appendix \ref{App:Tmatrix}, and we present the final result here:
\begin{align}
\label{eq:lim00} \lim\limits_{\bs{r} \to \bs{0}} I_{00} &= \frac{1}{4\pi \gamma_{01} i\omega }\left(\Omega_+ - \Omega_- \right),
 \\
\label{eq:lim03} \lim\limits_{\bs{r} \to \bs{0}} I_{03} &=  -\frac{1}{4\pi \gamma_{01} i\omega}\sum\limits_{\sigma = \pm} \sigma \left[ \gamma_\mathcal{E} \Omega_\sigma^2 + \Omega_\sigma^2 \ln \frac{\Omega_\sigma a}{2}\right],
\end{align}
where $\gamma_\mathcal{E}$ is the Euler-Mascheroni constant, $\Omega_\sigma \equiv \sqrt{i\omega (\pm \gamma_{01} - i\omega)}$, and the lattice constant $a$ is used as an infrared cutoff. Substituting Eqs.~(\ref{eq:lim00}) and (\ref{eq:lim03}) into Eqs.~(\ref{eq:f}) and (\ref{eq:g}) we obtain the final analytical form for $f(i\omega)$ and $g(i\omega)$.

\subsection{Local Density of States}

To find the first-order correction in $\gamma_{03}$ to the local density of states we use Eq.~(\ref{Grealspace}) and we proceed as follows:
\begin{align}
\nonumber \delta\rho(i\omega,\bs{r}) = - \frac{1}{\pi} \im \tr_1 \left[G(i\omega,\bs{r}) T(i\omega) G(i\omega,-\bs{r})\right] \approx \phantom{aaaaaa}\\
\nonumber  -\frac{1}{\pi} \im \tr_1 \left[G_0(i\omega,\bs{r}) T(i\omega) G_0(i\omega,-\bs{r}) + \right. \phantom{aaaaa} \\
\label{eq:LDOS}\left. G_0(i\omega,\bs{r}) T(i\omega) G_1(i\omega,-\bs{r})  + G_1(i\omega,\bs{r}) T(i\omega) G_0(i\omega,-\bs{r})\right],
\end{align}
where we only take the trace of the \textit{first-layer} components of the matrix, i.e., $\tr_1 M = M_{11} + M_{22}$. We choose to calculate only a partial trace because the scanning tunneling microscope tip measures mostly the electronic density of states of the topmost layer. Since we are dealing with a perturbative calculation in $\gamma_{03}$ up to the first order, we omitted the term $G_1(i\omega,\bs{r}) T(i\omega) G_1(i\omega,-\bs{r})$ proportional to $\gamma_{03}^2$. Note also that $G_0(i\omega,\bs{r})$ and $G_1(i\omega,\bs{r})$ were calculated in polar coordinates, thus replacing $\bs{r} \to -\bs{r}$ is equivalent to $\phi_r \to \pi + \phi_r$. We also keep in mind that to calculate the physical response we should use the retarded Green's functions, in other words, we should replace $i\omega \to \epsilon + i \delta$, where $\epsilon$ is the energy and $\delta \to +0$ is a positive infinitesimal shift. 

\subsubsection{Real space}

\begin{figure}
\centering
	\includegraphics[width = 0.485\columnwidth]{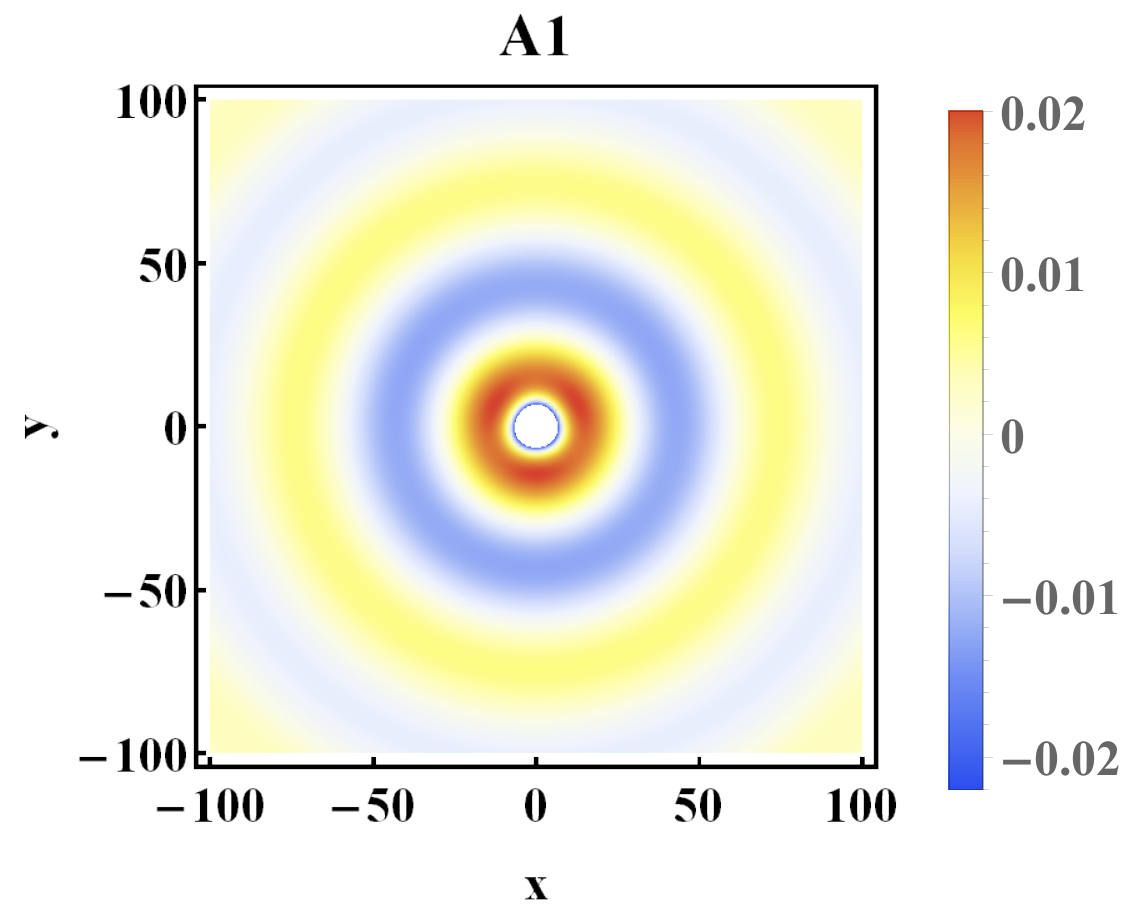}
	\includegraphics[width = 0.495\columnwidth]{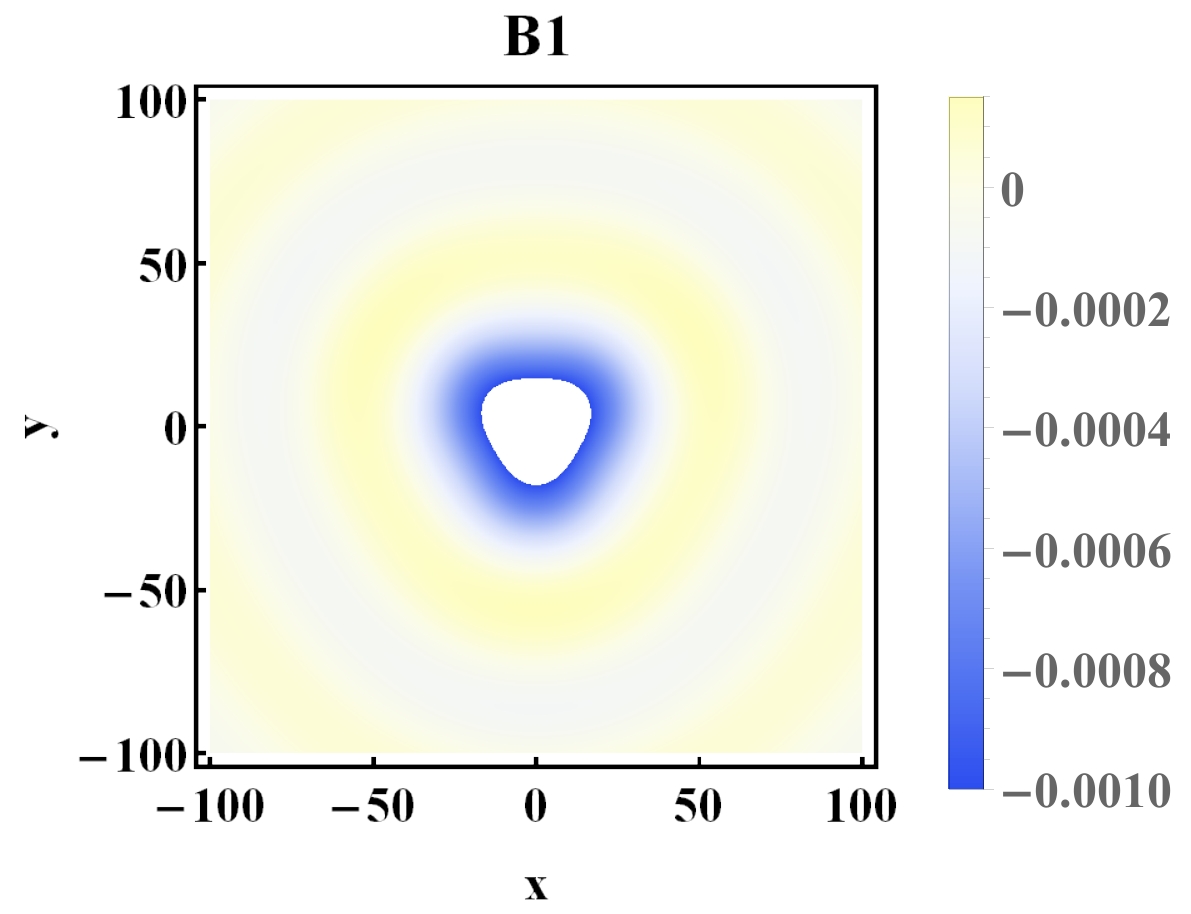}\\
	\includegraphics[width = 0.49\columnwidth]{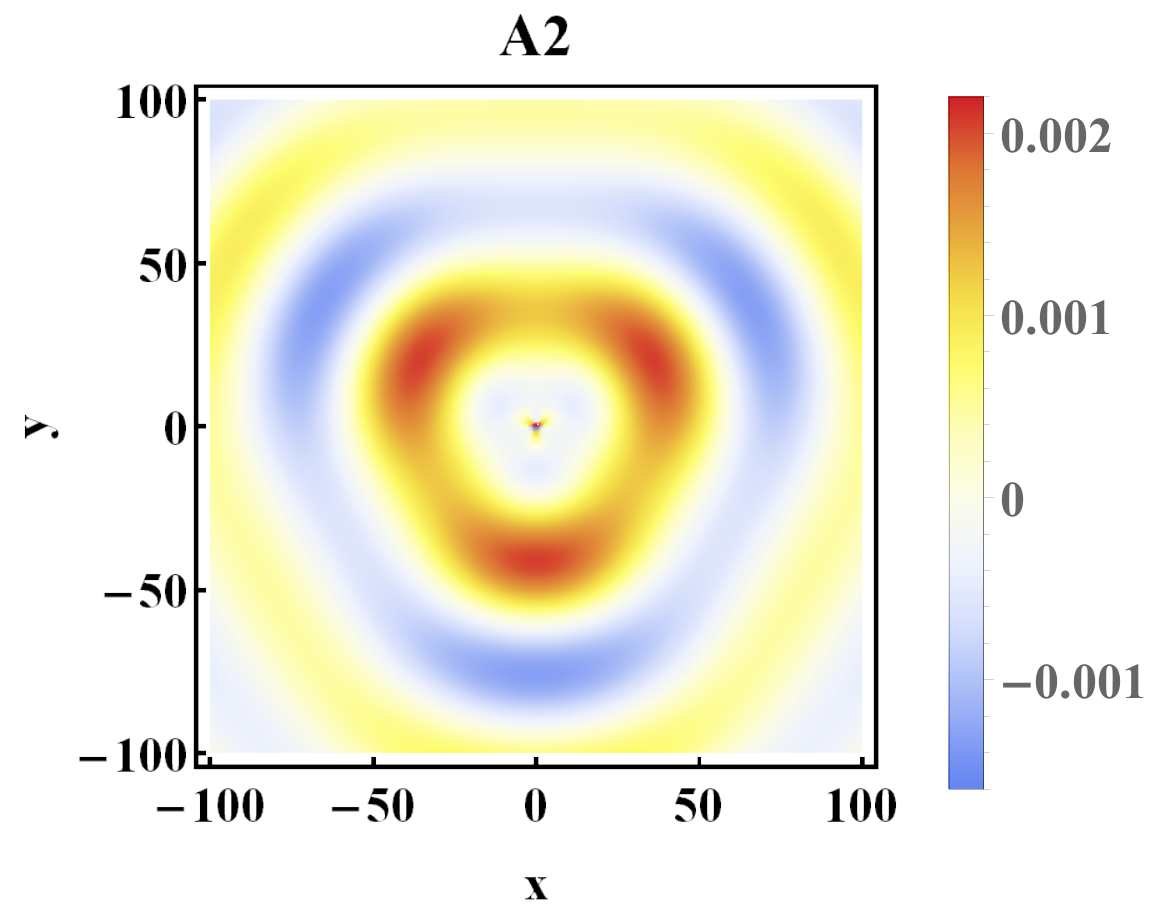}
	\includegraphics[width = 0.49\columnwidth]{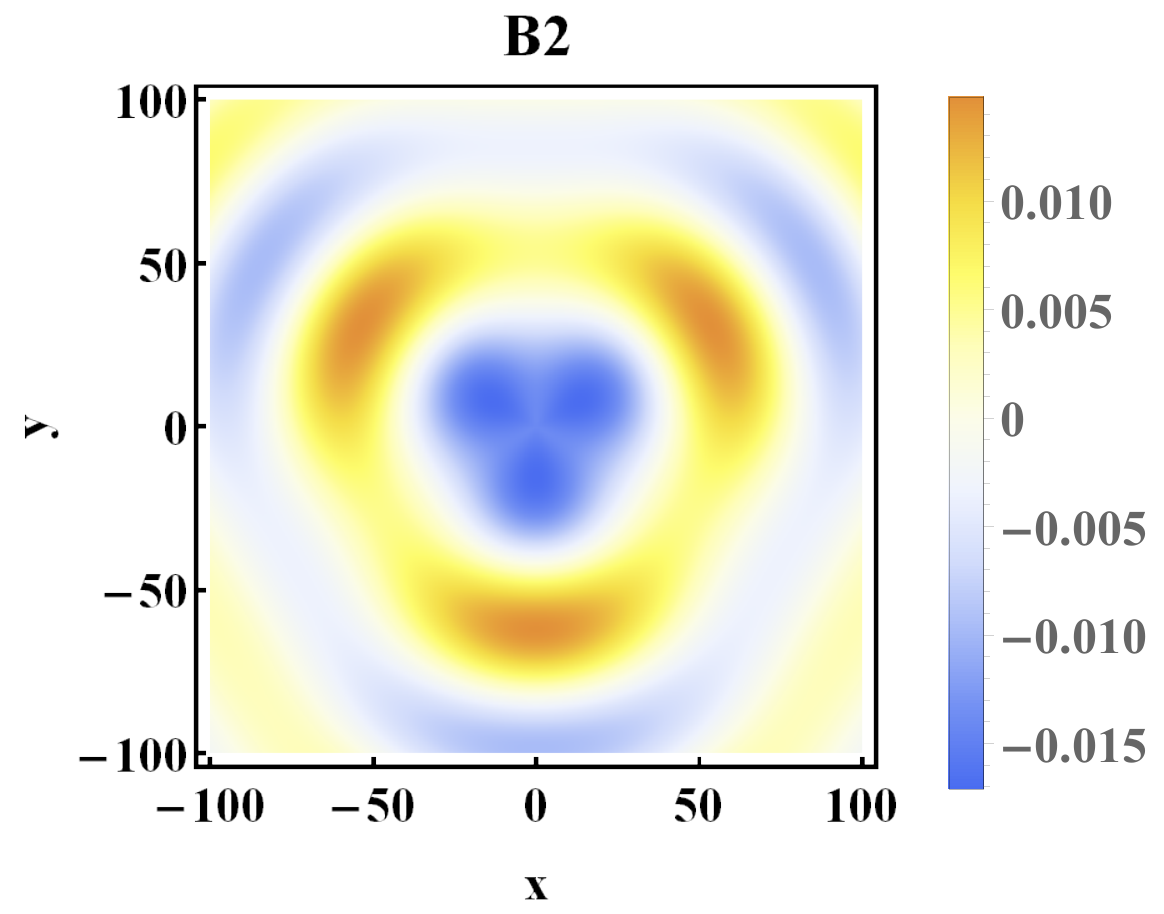}\\	
	\caption{Corrections to the local density of states calculated for $A1$, $B1$, $A2$, and $B2$ impurities, plotted as a function of $x$ and $y$ taken in the units of interatomic distance $a^* = a / \sqrt{3} = 1.42\,$\AA. We set $\gamma_{01} \approx 0.15$, $\gamma_{03} \approx - 0.09$, $U \approx - 105$, $\epsilon \approx 0.018$, which in dimensionful units corresponds to $\gamma_{1} = 0.42\,$eV, $\gamma_{3} = -3.3\,$eV, $U = -300\,$eV, $E = 50\,$meV.}
	\label{fig_A1_B1_A2_B2_RS}
\end{figure}

Considering the form of the $T$-matrix presented in Eq.~(\ref{eq:Tmatrices}), we can calculate analytically via Eq.~(\ref{eq:LDOS}) the corrections to the local density of states induced by each type of impuritiy. The exact analytical expressions for $\delta\rho_{A1}(\epsilon,r,\phi_r)$, $\delta\rho_{B1}(\epsilon,r,\phi_r)$, $\delta\rho_{A2}(\epsilon,r,\phi_r)$, and $\delta\rho_{B2}(\epsilon,r,\phi_r)$ can be found using Eq.~\ref{eq:LDOS} and the Appendixes. We plot the corresponding expressions in Fig.~\ref{fig_A1_B1_A2_B2_RS}. 

We can clearly see that the figures for A2 and B2 impurities show strong threefold-symmetric features originating from the trigonal warping terms. Using the asymptotic forms of the local density of states calculated at $r \to + \infty$ from Eqs.~(\ref{App:eq:rhoA1}-\ref{App:eq:rhoB2}), we present below their simplified forms at energies such that $0 < \epsilon \ll \gamma_{01}$. In this range of energies we can neglect the exponentially fast decaying terms, i.e., the terms containing $e^{-2\Omega_+ r}$ factors, and thus we get:
\begin{align}
\label{eq:rhoA1} \delta\rho_{A1} = \frac{1}{\pi} \im \left[ \frac{f(\epsilon)}{32 \pi \epsilon} \Omega_- \frac{\gamma_{01} + \gamma_{03} \Omega_- \sin 3\phi_r}{r} e^{-2\Omega_- r}\right]
\end{align}
\begin{align}
\label{eq:rhoB1} \delta\rho_{B1} = \frac{1}{\pi} \im \left[ \frac{g(\epsilon)}{32 \pi} \gamma_{01} \frac{\epsilon  + \gamma_{03} \Omega_- \sin 3\phi_r}{\Omega_- r} e^{-2\Omega_- r}\right] 
\end{align}
\begin{align}
\label{eq:rhoA2} \delta\rho_{A2} =\frac{1}{\pi} \im \left[ \frac{g(\epsilon)}{32 \pi} \gamma_{01} \frac{\epsilon - \gamma_{03} \Omega_- \sin 3\phi_r}{\Omega_- r} e^{-2\Omega_- r}\right] 
\end{align}
\begin{align}
\label{eq:rhoB2}  \delta\rho_{B2} = - \frac{1}{\pi} \im \left[ \frac{f(\epsilon)}{32 \pi} \gamma_{01} \frac{\gamma_{01} - \gamma_{03} \Omega_- \sin 3\phi_r}{ \Omega_- r} e^{-2\Omega_- r}\right]
\end{align}
Note, that for negative energies $\epsilon < 0$ we should keep the terms with $e^{-2\Omega_+ r}$ factors and discard those with $e^{-2\Omega_- r}$. Furthermore, it is worth emphasizing that the asymptotic expansions above describe the LDOS well only at large distances from the origin, namely, at $r \gg \mathrm{max}(\frac{1}{2|\Omega_+|},\frac{1}{2|\Omega_-|})$. Therefore, to recover correctly the features in the vicinity of the impurity we have to consider the full expressions provided in Eqs.~(\ref{App:eq:rhoA1}-\ref{App:eq:rhoB2}). 

The results presented in Fig.~\ref{fig_A1_B1_A2_B2_RS} are in a nearly perfect agreement with the semi-analytical $T$-matrix calculations based on a lattice model and presented in Ref.~[\onlinecite{Joucken2021}].

Furthermore, the orientation of the triangles flips with the sign of $\gamma_{03}$. This can be straightforwardly seen in the asymptotic forms of the corrections derived in Eqs.~(\ref{eq:rhoA1}-\ref{eq:rhoB2}) for $r \to +\infty$, that can be rewritten in a simplified form
\begin{align}
\delta\rho \sim \frac{\alpha + \beta \gamma_{03} \sin 3\phi_r}{r} e^{-\Omega r},
\label{eq:deltarhoall}
\end{align}
where $\alpha, \beta$ and $\Omega$ can be inferred from Eqs.~(\ref{eq:rhoA1}-\ref{eq:rhoB2}). 

In the equation above the term generated by trigonal warping is proportional to $\sin 3\phi_r/r$, while the first term proportional to $1/r$ is known in the literature as the quasiparticle interference pattern for bilayer graphene in the absence of trigonal warping \cite{Bena2008}. 

We should note, as expected from the band structure, the smaller the energy at which we calculate the quasiparticle interference patterns, the more visible are the effects of the trigonal warping.

\subsubsection{Momentum space}

\begin{figure}[h!]
	\centering
	\includegraphics[width = 0.49\columnwidth]{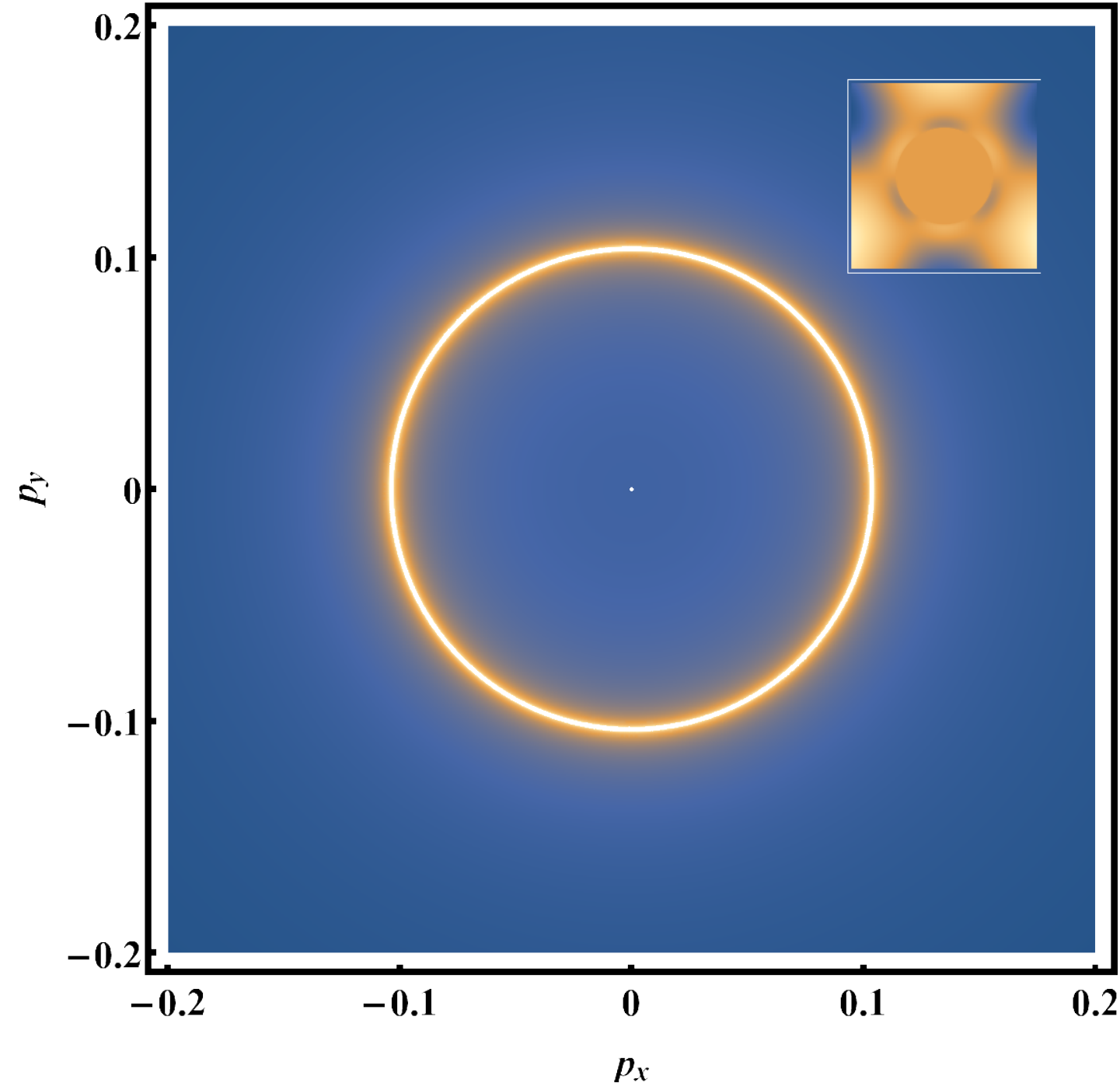}
	\includegraphics[width = 0.49\columnwidth]{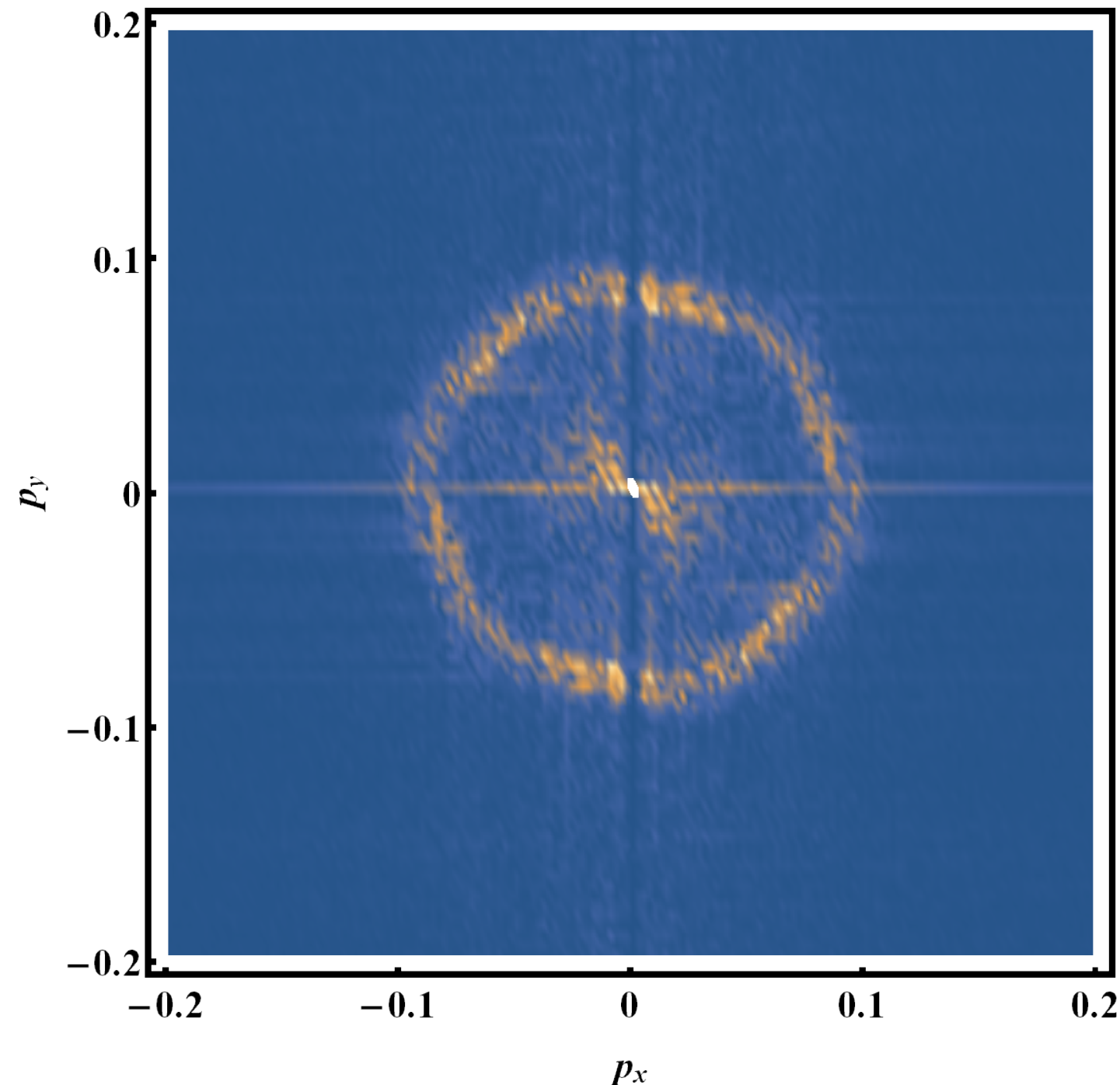}
	\caption{Analytically calculated and experimentally measured quasiparticle interference patterns are presented in dimensionless momentum space in the left and right panels, respectively.  The analytical panel and its inset are obtained as the absolute value and the imaginary part (normalized by the response without trigonal warping) of the sum of the responses to different types of impurities given in Eqs.~(\ref{eq:rhoA1FT}-\ref{eq:rhoB2FT}). The inset demonstrates the hexagonal symmetry of the problem. The right panel shows a fast Fourier transform of an experimental scanning tunneling microscopy $dI/dV$ map obtained on Bernal-stacked bilayer graphene, reproduced from Fig.~2g of Ref.~[\onlinecite{Joucken2021}] (see Ref.~[\onlinecite{Joucken2021}] for experimental details). It was acquired at $E = 47\,$meV, within the valence band. Both panels show a ring-like resonance appearing at $p_{\mathrm{res}} \approx 0.1$ given by Eq.~(\ref{eq:presvalue}) and defined by the energy at which the local density of states is calculated, as well as by the intra- and interlayer coupling constants.  We take the same values of parameters as in Fig.~2g in Ref.~[\onlinecite{Joucken2021}]: $\gamma_0 = 3.3\,$eV, $\gamma_{1} = 0.42\,$eV, $\gamma_{3} = -0.3\,$eV, $E = 47\,$meV. Additionally, we set $U = -300\,$eV, while in Ref.~[\onlinecite{Joucken2021}] $U = -10\,$eV.}
	\label{fig_A1_B1_A2_B2_MS}
\end{figure}

In what follows we analyze theoretically momentum-space quasiparticle interference patterns experimentally accessible via Fourier-transform scanning tunneling microscopy. It is clear that the Fourier transform of Eq.~(\ref{eq:deltarhoall}) yields both rotationally symmetric and angular-dependent parts. We Fourier-transform Eqs.~(\ref{eq:rhoA1}-\ref{eq:rhoB2}) and we get: 
\begin{widetext}
\begin{align}
\label{eq:rhoA1FT} \delta\rho_{A1} &= \frac{1}{2\pi i} \frac{ f(\epsilon) \Omega_- \left[ \gamma_{01} \mathcal{F}_0(p,2\Omega_-) + \gamma_{03} \Omega_- \mathcal{F}_1 (\bs{p}, 2\Omega_-) \right] -  f^*(\epsilon) \Omega^*_- \left[ \gamma_{01} \mathcal{F}_0(p,2\Omega^*_-) + \gamma_{03} \Omega^*_- \mathcal{F}_1(\bs{p}, 2\Omega^*_-) \right]}{32 \pi \epsilon}, \\
\label{eq:rhoB1FT} \delta\rho_{B1} &= \frac{1}{2\pi i} \gamma_{01} \frac{ \frac{g(\epsilon)}{ \Omega_-} \left[ \epsilon \mathcal{F}_0(p,2\Omega_-) + \gamma_{03} \Omega_- \mathcal{F}_1 (\bs{p}, 2\Omega_-) \right] -  \frac{g^*(\epsilon)}{ \Omega^*_-} \left[ \epsilon \mathcal{F}_0(p,2\Omega^*_-) + \gamma_{03} \Omega^*_- \mathcal{F}_1(\bs{p}, 2\Omega^*_-) \right]}{32 \pi}, \\
\label{eq:rhoA2FT} \delta\rho_{A2} &= \frac{1}{2\pi i} \gamma_{01} \frac{ \frac{g(\epsilon)}{ \Omega_-} \left[ \epsilon \mathcal{F}_0(p,2\Omega_-) - \gamma_{03} \Omega_- \mathcal{F}_1 (\bs{p}, 2\Omega_-) \right] -  \frac{g^*(\epsilon)}{ \Omega^*_-} \left[ \epsilon \mathcal{F}_0(p,2\Omega^*_-) - \gamma_{03} \Omega^*_- \mathcal{F}_1(\bs{p}, 2\Omega^*_-) \right]}{32 \pi}, \\
\label{eq:rhoB2FT}  \delta\rho_{B2} &= - \frac{1}{2\pi i} \gamma_{01} \frac{ \frac{f(\epsilon)}{ \Omega_-} \left[ \gamma_{01} \mathcal{F}_0(p,2\Omega_-) - \gamma_{03} \Omega_- \mathcal{F}_1 (\bs{p}, 2\Omega_-) \right] -  \frac{f^*(\epsilon)}{ \Omega^*_-} \left[ \gamma_{01} \mathcal{F}_0(p,2\Omega^*_-) - \gamma_{03} \Omega^*_- \mathcal{F}_1(\bs{p}, 2\Omega^*_-) \right]}{32 \pi},
\end{align}
\end{widetext}
where $\mathcal{F}_0$ and $\mathcal{F}_1$  are calculated in Appendix \ref{App:LDOS}, and are given by:
\begin{align}
&\mathcal{F}_0(p,\Omega) = \frac{2\pi}{\Omega} \frac{1}{\sqrt{1+\frac{p^2}{\Omega^2}}} \\
\nonumber &\mathcal{F}_1(\bs{p},\Omega) = 2\pi i \sin 3\phi_p \times \\
&\frac{p^2\left(-3+ \sqrt{1+\frac{p^2}{\Omega^2}} \right) + 4\Omega^2\left(-1+ \sqrt{1+\frac{p^2}{\Omega^2}} \right)}{p^3\sqrt{1+\frac{p^2}{\Omega^2}}}
\end{align} 
Using these definitions and the Fourier transforms in Eqs.~(\ref{eq:rhoA1FT}-\ref{eq:rhoB2FT}) we infer the position of the ring-like resonance in momentum space
\begin{align}
p_{\mathrm{res}} = - 2 i \Omega_- =  2\sqrt{\epsilon(\gamma_{01} + \epsilon)}.
\label{eq:pres}
\end{align}
Above we denoted $\epsilon \equiv \frac{2E}{\gamma_0 \sqrt{3}}$, where $E$ is the energy at which we calculate the response. On the left panel in Fig.~\ref{fig_A1_B1_A2_B2_MS} we plot the absolute of the sum of Fourier transforms in Eqs.~(\ref{eq:rhoA1FT}-\ref{eq:rhoB2FT}) simulating a large-sample experiment in which all types of impurities contribute to the local response. On the right panel of Fig.~\ref{fig_A1_B1_A2_B2_MS} we show the corresponding experimental measurement in dimensionless momentum space $(p_x, p_y) = (k_x a, k_y a)$, where the lattice constant $a = 2.46\,$\AA. We see that the analytical and experimental results are in a good agreement, i.e., $p^{\mathrm{exp}}_{\mathrm{res}} \approx 0.095$ is very close to the analytically calculated value using Eq.~(\ref{eq:pres}): 
\begin{align}
p_{\mathrm{res}} = 2 \sqrt{\frac{2E}{\gamma_0 \sqrt{3}}\left(\frac{2\gamma_1}{\gamma_0 \sqrt{3}}+  \frac{2E}{\gamma_0 \sqrt{3}}\right)} \approx 0.1,
\label{eq:presvalue}
\end{align}
where we set $E = 47\,$meV.  

To derive the FT LDOS we used the asymptotic expansions in Eqs.~(\ref{eq:rhoA1}-\ref{eq:rhoB2}), and since we work with asymptotic expansions at $r \to + \infty$  obtained within a low-energy approximation, the hexagonal shape of the resonance is lost in the analytically computed FT versus the experimental one (left versus right panels of Fig.~\ref{fig_A1_B1_A2_B2_MS}). In Eqs.~(\ref{eq:rhoA1FT}-\ref{eq:rhoB2FT}) the reminiscence of hexagonal symmetries is carried solely by the phase factor $\sin 3\phi_p$ in the definition of $\mathcal{F}_1(\bs{p},\Omega)$, and it is reflected in the inset of the left panel.

The results obtained above are also qualitatively consistent with the numerical $T$-matrix calculations presented in Ref.~[\onlinecite{Joucken2021}], with the only difference being the hexagonal shape of the resonance in the latter. This discrepancy stems from the fact that in this work we use a low-energy approximation, while the numerical $T$-matrix calculations were performed within a lattice model. 

\section{Conclusions}
We have calculated analytically the form of the quasiparticle interference patterns in bilayer graphene for four types of impurities localized on different layers and sublattices, taking into account perturbatively the trigonal warping of the bands. First and foremost, our results both in real space and in Fourier space are in good agreement with the experimental measurements and $T$-matrix-based semi-analytical simulations of such patterns \cite{Joucken2020,Joucken2021}. Most importantly the fact that our analytical results can be expressed in closed form provides us with an understanding of the origin of the observed triangular features in real space. Thus, we clearly see that they originate in the trigonal warping terms and flip orientation when the trigonal warping is changing sign. Also our results allow us to predict the value of the momentum corresponding to the ring-like resonances visible in momentum space and to the real-space oscillations; this seems to be independent of the value of the trigonal warping.

\acknowledgments

J.V.J. acknowledges support from the National Science Foundation under award DMR-1753367 and the Army Research Office under contract W911NF-17-1-0473.

\newpage
\bibliography{biblio_BLG}


\newpage
\widetext
\appendix

\section{Green's functions in momentum space}\label{App:GFMS}

\subsection{0-th order}\label{App:ZerothOrder}

\begin{align}
G_0(i\omega, \bs{q}) =
	\frac{1}{(q^2-(i\omega)^2+i\omega \gamma_{01} )(q^2-(i\omega)^2-i\omega\gamma_{01})} 
	\bpm
		g_0^{11} & \left(g_0^{21}\right)^\star & \left(g_0^{31}\right)^\star & \left(g_0^{41}\right)^\star \\
		g_0^{21} & g_0^{22} & g_0^{32} & \left(g_0^{31}\right)^\star \\
		g_0^{31} & g_0^{32} & g_0^{22} & \left(g_0^{21}\right)^\star \\
		g_0^{41} & g_0^{31} & g_0^{21} & g_0^{11}
	\epm, 
\end{align}
with $g_0^{ij}$ denoting the $(ij)$-th element of the matrix $g_0$.
\begin{align}
\nonumber &g_0^{11} \equiv i\omega((i\omega)^2 - \gamma_{01}^2) - i\omega q^2,\;
g_0^{22} \equiv (i\omega)^3 - (i\omega) q^2,\;
g_0^{21} \equiv (i\omega)^2 \cdot sqe^{is\phi_q} - s q^3 e^{is\phi_q}, \;
g_0^{31} \equiv  i\omega \gamma_{01} \cdot sqe^{is\phi_q},\\
&g_0^{32} \equiv (i\omega)^2 \gamma_{01},\; 
g_0^{41} \equiv \gamma_{01} \cdot q^2 e^{2is\phi_q}.
\end{align}
The symbol $\star$ denotes replacing $\phi_q \to -\phi_q$, e.g., $\left(g_0^{31}\right)^\star = i\omega \gamma_{01} \cdot sqe^{-is\phi_q}$. 

\subsection{1-st order}\label{App:FirstOrder}
The first-order correction in trigonal warping is given by
\begin{align}
\nonumber G_1(i\omega, \bs{q}) = 
	\frac{\gamma_{03}}{(q^2-(i\omega)^2+i\omega\gamma_{01})^2(q^2-(i\omega)^2-i\omega\gamma_{01})^2} 
	\bpm
		g_1^{11} & \left(g_1^{21}\right)^\star & \left(g_1^{31}\right)^\star & \left(g_1^{41}\right)^\star \\
		g_1^{21} & g_1^{22} & \left(g_1^{32}\right)^\star & \left(g_1^{31}\right)^\star \\
		g_1^{31} & g_1^{32} & g_1^{22} & \left(g_1^{21}\right)^\star \\
		g_1^{41} & g_1^{31} & g_1^{21} & g_1^{11}
	\epm, 
\end{align}
where
\begin{align}
g_1^{11} &\equiv - \gamma_{01} i\omega \left[ s q^5 e^{3 i s \phi_q} +  s q^5 e^{-3 i s \phi_q} \right] - \gamma_{01} i\omega (\gamma_{01}^2 - (i\omega)^2) \left[ s q^3 e^{3 i s \phi_q} +  s q^3 e^{-3 i s \phi_q} \right] \\
g_1^{22} &\equiv - \gamma_{01} i\omega \left[ s q^5 e^{3 i s \phi_q} +  s q^5 e^{-3 i s \phi_q} \right] + \gamma_{01} (i\omega)^3 \left[ s q^3 e^{3 i s \phi_q} +  s q^3 e^{-3 i s \phi_q} \right], \\
g_1^{21} &\equiv - \gamma_{01}\left[ q^6 e^{4 i s \phi_q} \right] + \gamma_{01}(i\omega)^2 \left[ q^4 e^{4 i s \phi_q} - q^4 e^{-2 i s \phi_q} \right] - \gamma_{01}(i\omega)^2 (\gamma_{01}^2 - (i\omega)^2) \left[ q^2 e^{-2 i s \phi_q} \right], \\
g_1^{31} &\equiv i\omega \left[ q^6 e^{-2 i s \phi_q} \right] + \gamma_{01}^2 i\omega\left[ q^4 e^{4 i s \phi_q} \right]  + i\omega (\gamma_{01}^2 - 2(i\omega)^2) \left[ q^4 e^{-2 i s \phi_q} \right] -  (i\omega)^3 (\gamma_{01}^2 - (i\omega)^2) \left[ q^2 e^{-2 i s \phi_q}\right], \\
g_1^{32} &\equiv \left[ s q^7 e^{-3 i s \phi_q} \right] - 2(i\omega)^2 \left[ sq^5 e^{-3is\phi_q} \right] + (i\omega)^4 \left[ sq^3 e^{-3is\phi_q} \right] + \gamma_{01}^2 (i\omega)^2 \left[ sq^3 e^{3is\phi_q} \right],\\
g_1^{41} &\equiv \gamma_{01}^2 \left[ sq^5 e^{5is\phi_q}  \right] + (i\omega)^2 \left[ sq^5 e^{-is\phi_q}  \right] + 2(i\omega)^2 (\gamma_{01}^2 - (i\omega)^2) \left[ sq^3 e^{-is\phi_q} \right] + (i\omega)^2 (\gamma_{01}^2 - (i\omega)^2)^2 \left[ sq e^{-is\phi_q} \right]. 
\end{align}

\section{Integrals to define Green's functions in real space}\label{App:IntegralsRS}

\subsection{Integrals for the 0-th order}\label{App:IntegralsZerothOrder}

We can write the 0-th order Green's function in real space as follows:
\begin{align}
\label{G0RS}
G_0(i\omega,\bs{r}) =
	\bpm
		i\omega((i\omega)^2-\gamma_{01}^2) I_{00} - i\omega I_{03} & (i\omega)^2 I^-_{01} - I_{04}^- & \gamma_{01} i\omega I_{01}^- & \gamma_{01} I_{02}^- \\
		(i\omega)^2 I^+_{01} - I_{04}^+ & i\omega^3 I_{00} - i\omega I_{03} & \gamma_{01} (i\omega)^2 I_{00} & \gamma_{01} i\omega I_{01}^- \\
		\gamma_{01} i\omega I_{01}^+ & \gamma_{01} (i\omega)^2 I_{00} & i\omega^3 I_{00} - i\omega I_{03} & (i\omega)^2 I^-_{01} - I_{04}^- \\
		\gamma_{01} I_{02}^+ & \gamma_{01}i\omega I_{01}^+ & (i\omega)^2 I^+_{01} - I_{04}^+ & i\omega((i\omega)^2-\gamma_{01}^2) I_{00} - i\omega I_{03}
	\epm
\end{align}

\begin{align}
\nonumber I_{00} &= \int\limits_0^{+\infty} \frac{qdq}{2\pi} \int\limits_0^{2\pi} \frac{d\phi_q}{2\pi} 	\frac{e^{i q r \cos(\phi_q - \phi_r)}}{(q^2-(i\omega)^2+\gamma_{01}i\omega)(q^2-(i\omega)^2-\gamma_{01}i\omega)}  = \int\limits_0^{+\infty} \frac{dq}{2\pi} \frac{q J_0(q r)}{(q^2-(i\omega)^2+\gamma_{01}i\omega)(q^2-(i\omega)^2-\gamma_{01}i\omega)} = \\
&= -\frac{1}{4\pi \gamma_{01} i\omega } \left[K_0(\Omega_+ r) - K_0(\Omega_- r)\right] \\
\nonumber I^\pm_{01} &= \int\limits_0^{+\infty} \frac{qdq}{2\pi} \int\limits_0^{2\pi} \frac{d\phi_q}{2\pi} 	\frac{s q e^{\pm i s \phi_q} e^{i q r \cos(\phi_q - \phi_r)}}{(q^2-(i\omega)^2+\gamma_{01}i\omega)(q^2-(i\omega)^2-\gamma_{01}i\omega)}  = \int\limits_0^{+\infty} \frac{dq}{2\pi} \frac{i s e^{\pm i s \phi_r} q^2 J_1(q r)}{(q^2-(i\omega)^2+\gamma_{01}i\omega)(q^2-(i\omega)^2-\gamma_{01}i\omega)} = \\
 &= -\frac{i s e^{\pm i s \phi_r}}{4\pi \gamma_{01} i \omega_n } \left[\Omega_+ K_1(\Omega_+ r) - \Omega_- K_1(\Omega_- r)\right] \\
\nonumber I^\pm_{02} &= \int\limits_0^{+\infty} \frac{qdq}{2\pi} \int\limits_0^{2\pi} \frac{d\phi_q}{2\pi} 	\frac{q^2 e^{\pm 2 i s \phi_q} e^{i q r \cos(\phi_q - \phi_r)}}{(q^2-(i\omega)^2+\gamma_{01}i\omega)(q^2-(i\omega)^2-\gamma_{01}i\omega)}  = \int\limits_0^{+\infty} \frac{dq}{2\pi} \frac{-e^{\pm 2i s \phi_r} q^3 J_2(q r)}{(q^2-(i\omega)^2+\gamma_{01}i\omega)(q^2-(i\omega)^2-\gamma_{01}i\omega)} = \\
 &= \frac{e^{\pm 2i s \phi_r}}{4\pi \gamma_{01} i\omega} \left[\Omega_+^2 K_2(\Omega_+ r) - \Omega_-^2 K_2(\Omega_- r)\right] \\
\nonumber I_{03} &= \int\limits_0^{+\infty} \frac{qdq}{2\pi} \int\limits_0^{2\pi} \frac{d\phi_q}{2\pi} 	\frac{q^2 e^{i q r \cos(\phi_q - \phi_r)}}{(q^2-(i\omega)^2+\gamma_{01}i\omega)(q^2-(i\omega)^2-\gamma_{01}i\omega)}  = \int\limits_0^{+\infty} \frac{dq}{2\pi} \frac{q^3 J_0(q r)}{(q^2-(i\omega)^2+\gamma_{01}i\omega)(q^2-(i\omega)^2-\gamma_{01}i\omega)} = \\
 &= \frac{1}{4\pi \gamma_{01} i\omega} \left[\Omega_+^2 K_0(\Omega_+ r) - \Omega_-^2 K_0(\Omega_- r)\right] \\
\nonumber I^\pm_{04} &= \int\limits_0^{+\infty} \frac{qdq}{2\pi} \int\limits_0^{2\pi} \frac{d\phi_q}{2\pi} 	\frac{s q^3 e^{\pm i s \phi_q} e^{i q r \cos(\phi_q - \phi_r)}}{(q^2-(i\omega)^2+\gamma_{01}i\omega)(q^2-(i\omega)^2-\gamma_{01}i\omega)}  = \int\limits_0^{+\infty} \frac{dq}{2\pi} \frac{i s e^{\pm i s \phi_r} q^4 J_1(q r)}{(q^2-(i\omega)^2+\gamma_{01}i\omega)(q^2-(i\omega)^2-\gamma_{01}i\omega)} = \\
 &= \frac{i s e^{\pm i s \phi_r}}{4\pi \gamma_{01} i\omega } \left[\Omega_+^3 K_1(\Omega_+ r) - \Omega_-^3 K_1(\Omega_- r)\right],
\end{align}
where we denoted
$
\Omega_\pm = \sqrt{i\omega( \pm \gamma_{01} - i\omega)},
$
and $K_{i}(z)$ is the $i$-th modified Bessel function of the second kind.

\subsection{Integrals for the 1-st order}\label{App:IntegralsFirstOrder}
We can write component-wise the 1-st order Green's function in real space as follows:
\begin{align}
\label{G111}
G^{11}_1 &= \gamma_{03} \left[ -\gamma_{01}i\omega (I_{53}^++I_{53}^-) - \gamma_{01}i\omega(\gamma_{01}^2-(i\omega)^2)(I_{33}^++I_{33}^-) \right], \\
G^{22}_1 &= \gamma_{03} \left[ -\gamma_{01}i\omega (I_{53}^++I_{53}^-) + \gamma_{01}(i\omega)^3(I_{33}^++I_{33}^-) \right],\\
G^{21}_1 &= \gamma_{03} \left[-\gamma_{01} I_{64}^+ + \gamma_{01} (i\omega)^2 (I_{44}^+ -I_{42}^-) - \gamma_{01}(i\omega)^2 (\gamma_{01}^2-(i\omega)^2) I_{22}^- \right],\\
G^{31}_1 &= \gamma_{03} \left[ i\omega I_{62}^- +\gamma_{01}^2 i\omega I_{44}^+ + i\omega(\gamma_{01}^2-2(i\omega)^2)I_{42}^- - (i\omega)^3(\gamma_{01}^2-(i\omega)^2) I_{22}^- \right],\\
G^{32}_1 &= \gamma_{03} \left[ I_{73}^- - 2(i\omega)^2 I_{53}^- + (i\omega)^4 I_{33}^- + \gamma_{01}^2 (i\omega)^2 I_{33}^+ \right],\\
G^{41}_1 &= \gamma_{03} \left[ \gamma_{01}^2 I_{55}^+ + (i\omega)^2 I_{51}^- + 2(i\omega)^2 (\gamma_{01}^2-(i\omega)^2) I_{31}^- + (i\omega)^2 (\gamma_{01}^2-(i\omega)^2)^2I_{11}^- \right], \\
G^{33}_1 &= G^{22}_1,\, G^{44}_1 = G^{11}_1,\,G^{42}_1 = G^{31}_1,\, G^{43}_1 = G^{21}_1,  \\
\label{G121} G^{12}_1 &= G^{34}_1 = G^{21}_1( + \leftrightarrow -),\,
G^{13}_1 = G^{24}_1 = G^{31}_1( + \leftrightarrow -),\,
G^{14}_1 = G^{41}_1( + \leftrightarrow -),\,
G^{23}_1 = G^{32}_1( + \leftrightarrow -).
\end{align}

\begin{align}
\nonumber I^\pm_{11} &= \int\limits_0^{+\infty} \frac{qdq}{2\pi} \int\limits_0^{2\pi} \frac{d\phi_q}{2\pi} 	\frac{s q e^{\pm i s \phi_q} e^{i q r \cos(\phi_q - \phi_r)}}{(q^2-(i\omega)^2+\gamma_{01}i\omega)^2(q^2-(i\omega)^2-\gamma_{01}i\omega)^2}  = \int\limits_0^{+\infty} \frac{dq}{2\pi} \frac{i s e^{\pm i s \phi_r} q^2 J_1(q r)}{(q^2-(i\omega)^2+\gamma_{01}i\omega)^2(q^2-(i\omega)^2-\gamma_{01}i\omega)^2} = \\
&= \frac{i s e^{\pm i s \phi_r}}{16\pi \gamma_{01}^3 (i\omega)^{3}} \left[2 \left( \Omega_+ K_1(\Omega_+ r) - \Omega_- K_1(\Omega_- r) \right) + i\omega \gamma_{01} r\left( K_0(\Omega_+ r) + K_0(\Omega_- r)\right) \right] \\
\nonumber I^\pm_{31} &= \int\limits_0^{+\infty} \frac{qdq}{2\pi} \int\limits_0^{2\pi} \frac{d\phi_q}{2\pi} 	\frac{s q^3 e^{\pm i s \phi_q} e^{i q r \cos(\phi_q - \phi_r)}}{(q^2-(i\omega)^2+\gamma_{01}i\omega)^2(q^2-(i\omega)^2-\gamma_{01}i\omega)^2}  = \int\limits_0^{+\infty} \frac{dq}{2\pi} \frac{i s e^{\pm i s \phi_r} q^4 J_1(q r)}{(q^2-(i\omega)^2+\gamma_{01}i\omega)^2(q^2-(i\omega)^2-\gamma_{01}i\omega)^2} = \\
&= \frac{i s e^{\pm i s \phi_r}}{16\pi \gamma_{01}^3 (i\omega)^2} \left[ 2 i\omega \left(\Omega_+ K_1(\Omega_+ r) - \Omega_- K_1(\Omega_- r) \right) - \gamma_{01} r \left(\Omega_+^2 K_0(\Omega_+ r) + \Omega_-^2 K_0(\Omega_- r) \right)\right]\\
\nonumber I^\pm_{51} &= \int\limits_0^{+\infty} \frac{qdq}{2\pi} \int\limits_0^{2\pi} \frac{d\phi_q}{2\pi} 	\frac{s q^5 e^{\pm i s \phi_q} e^{i q r \cos(\phi_q - \phi_r)}}{(q^2-(i\omega)^2+\gamma_{01}i\omega)^2(q^2-(i\omega)^2-\gamma_{01}i\omega)^2}  = \int\limits_0^{+\infty} \frac{dq}{2\pi} \frac{i s e^{\pm i s \phi_r} q^6 J_1(q r)}{(q^2-(i\omega)^2+\gamma_{01}i\omega)^2(q^2-(i\omega)^2-\gamma_{01}i\omega)^2} = \\
&= \frac{i s e^{\pm i s \phi_r}}{16\pi \gamma_{01}^3 (i\omega)^2} \left[2i\omega (\gamma_{01}^2-(i\omega)^2)\left(\Omega_- K_1(\Omega_- r) - \Omega_+ K_1(\Omega_+ r) \right) + \gamma_{01}r \left(\Omega_+^4 K_0(\Omega_+ r) + \Omega_-^4 K_0(\Omega_- r) \right) \right] \\
\nonumber I^\pm_{22} &= \int\limits_0^{+\infty} \frac{qdq}{2\pi} \int\limits_0^{2\pi} \frac{d\phi_q}{2\pi} 	\frac{q^2 e^{\pm 2 i s \phi_q} e^{i q r \cos(\phi_q - \phi_r)}}{(q^2-(i\omega)^2+\gamma_{01}i\omega)^2(q^2-(i\omega)^2-\gamma_{01}i\omega)^2}  = \int\limits_0^{+\infty} \frac{dq}{2\pi} \frac{-e^{\pm 2i s \phi_r} q^3 J_2(q r)}{(q^2-(i\omega)^2+\gamma_{01}i\omega)^2(q^2-(i\omega)^2-\gamma_{01}i\omega)^2} = \\
&=-\frac{e^{\pm 2 i s \phi_r}}{16\pi \gamma_{01}^3 (i\omega)^3} \left[  2 \left(  \Omega_+^2 K_2(\Omega_+ r) - \Omega_-^2 K_2(\Omega_- r) \right) + i\omega \gamma_{01} r\left( \Omega_+ K_1(\Omega_+ r) + \Omega_- K_1(\Omega_- r)\right) \right] \\
\nonumber I^\pm_{42} &= \int\limits_0^{+\infty} \frac{qdq}{2\pi} \int\limits_0^{2\pi} \frac{d\phi_q}{2\pi} 	\frac{q^4 e^{\pm 2 i s \phi_q} e^{i q r \cos(\phi_q - \phi_r)}}{(q^2-(i\omega)^2+\gamma_{01}i\omega)^2(q^2-(i\omega)^2-\gamma_{01}i\omega)^2}  = \int\limits_0^{+\infty} \frac{dq}{2\pi} \frac{-e^{\pm 2i s \phi_r} q^5 J_2(q r)}{(q^2-(i\omega)^2+\gamma_{01}i\omega)^2(q^2-(i\omega)^2-\gamma_{01}i\omega)^2} = \\
&=-\frac{e^{\pm 2 i s \phi_r}}{16\pi \gamma_{01}^3 (i\omega)^2} \left[  2i\omega \left(  \Omega_+^2 K_2(\Omega_+ r) - \Omega_-^2 K_2(\Omega_- r) \right) - \gamma_{01} r\left( \Omega_+^3 K_1(\Omega_+ r) + \Omega_-^3 K_1(\Omega_- r)\right) \right] \\
\nonumber I^\pm_{62} &= \int\limits_0^{+\infty} \frac{qdq}{2\pi} \int\limits_0^{2\pi} \frac{d\phi_q}{2\pi} 	\frac{q^6 e^{\pm 2 i s \phi_q} e^{i q r \cos(\phi_q - \phi_r)}}{(q^2-(i\omega)^2+\gamma_{01}i\omega)^2(q^2-(i\omega)^2-\gamma_{01}i\omega)^2}  = \int\limits_0^{+\infty} \frac{dq}{2\pi} \frac{-e^{\pm 2i s \phi_r} q^7 J_2(q r)}{(q^2-(i\omega)^2+\gamma_{01}i\omega)^2(q^2-(i\omega)^2-\gamma_{01}i\omega)^2} = \\
&=\frac{e^{\pm 2 i s \phi_r}}{16\pi \gamma_{01}^3 (i\omega)^2} \left[  2i\omega (\gamma_{01}^2 - (i\omega)^2)\left(  \Omega_+^2 K_2(\Omega_+ r) - \Omega_-^2 K_2(\Omega_- r) \right) - \gamma_{01} r\left( \Omega_+^5 K_1(\Omega_+ r) + \Omega_-^5 K_1(\Omega_- r)\right) \right] \\
\nonumber I^\pm_{33} &= \int\limits_0^{+\infty} \frac{qdq}{2\pi} \int\limits_0^{2\pi} \frac{d\phi_q}{2\pi} 	\frac{s q^3 e^{\pm 3 i s \phi_q} e^{i q r \cos(\phi_q - \phi_r)}}{(q^2-(i\omega)^2+\gamma_{01}i\omega)^2(q^2-(i\omega)^2-\gamma_{01}i\omega)^2}  = \int\limits_0^{+\infty} \frac{dq}{2\pi} \frac{-is e^{\pm 3i s \phi_r}q^4 J_3(q r)}{(q^2-(i\omega)^2+\gamma_{01}i\omega)^2(q^2-(i\omega)^2-\gamma_{01}i\omega)^2} = \\
&= -\frac{i s e^{\pm 3 i s \phi_r}}{16\pi \gamma_{01}^3 (i\omega)^3} \left[ 2 \left( \Omega_+^3 K_3(\Omega_+ r) -\Omega_-^3 K_3(\Omega_- r)\right) + i\omega \gamma_{01} r\left( \Omega_+^2 K_2(\Omega_+ r) + \Omega_-^2 K_2(\Omega_- r)\right) \right] \\
\nonumber I^\pm_{53} &= \int\limits_0^{+\infty} \frac{qdq}{2\pi} \int\limits_0^{2\pi} \frac{d\phi_q}{2\pi} 	\frac{s q^5 e^{\pm 3 i s \phi_q} e^{i q r \cos(\phi_q - \phi_r)}}{(q^2-(i\omega)^2+\gamma_{01}i\omega)^2(q^2-(i\omega)^2-\gamma_{01}i\omega)^2}  = \int\limits_0^{+\infty} \frac{dq}{2\pi} \frac{-is e^{\pm 3i s \phi_r}q^6 J_3(q r)}{(q^2-(i\omega)^2+\gamma_{01}i\omega)^2(q^2-(i\omega)^2-\gamma_{01}i\omega)^2} = \\
&=  -\frac{i s e^{\pm 3 i s \phi_r}}{16\pi \gamma_{01}^3 (i\omega)^2}\left[2 i \omega_n  \left(\Omega_+^3 K_3(\Omega_+r )-\Omega_-^3 K_3(\Omega_- r)\right) -\gamma_{01} r \left(\Omega_+^4 K_2(\Omega_+ r )+\Omega_-^4 K_2(\Omega_-r )\right)\right] \\
\nonumber I^\pm_{73} &= \int\limits_0^{+\infty} \frac{qdq}{2\pi} \int\limits_0^{2\pi} \frac{d\phi_q}{2\pi} 	\frac{s q^7 e^{\pm 3 i s \phi_q} e^{i q r \cos(\phi_q - \phi_r)}}{(q^2-(i\omega)^2+\gamma_{01}i\omega)^2(q^2-(i\omega)^2-\gamma_{01}i\omega)^2}  = \int\limits_0^{+\infty} \frac{dq}{2\pi} \frac{-is e^{\pm 3i s \phi_r}q^8 J_3(q r)}{(q^2-(i\omega)^2+\gamma_{01}i\omega)^2(q^2-(i\omega)^2-\gamma_{01}i\omega)^2} = \\
\nonumber &= \frac{i s e^{\pm 3 i s \phi_r}}{16\pi \gamma_{01}^3 (i\omega)^2}\left[2 i \omega_n  \left(\Omega_+^5 K_1(\Omega_+r )-\Omega_-^5 K_1(\Omega_- r)\right) -\gamma_{01} r \left(\Omega_+^6 K_0(\Omega_+ r )+\Omega_-^6 K_0(\Omega_-r )\right)+\right. \\
&\phantom{aaaaaaaaaaaaaaaaaaaaaaaaaaaaaaaaaaaaaaaaaaaa}\left. + \frac{8i\omega(\gamma_{01}^2-(i\omega)^2)}{r} \left(\Omega_+^2 K_2(\Omega_+ r )-\Omega_-^2 K_2(\Omega_-r )\right)\right] 
\end{align}
\begin{align}
\nonumber I^\pm_{44} &= \int\limits_0^{+\infty} \frac{qdq}{2\pi} \int\limits_0^{2\pi} \frac{d\phi_q}{2\pi} 	\frac{q^4 e^{\pm 4i s \phi_q} e^{i q r \cos(\phi_q - \phi_r)}}{(q^2-(i\omega)^2+\gamma_{01}i\omega)^2(q^2-(i\omega)^2-\gamma_{01}i\omega)^2}  = \int\limits_0^{+\infty} \frac{dq}{2\pi} \frac{ e^{\pm 4 i s \phi_r} q^5 J_4(q r)}{(q^2-(i\omega)^2+\gamma_{01}i\omega)^2(q^2-(i\omega)^2-\gamma_{01}i\omega)^2} = \\
&= \frac{e^{\pm 4 i s \phi_r}}{16\pi \gamma_{01}^3 (i\omega)^3} \left[2\left(\Omega_+^4 K_4(\Omega_+ r)-\Omega_-^4 K_4(\Omega_- r) \right) + i\omega \gamma_{01} r \left(\Omega_+^3 K_3(\Omega_+ r)+\Omega_-^3 K_3(\Omega_- r) \right)\right] \\
\nonumber I^\pm_{64} &= \int\limits_0^{+\infty} \frac{qdq}{2\pi} \int\limits_0^{2\pi} \frac{d\phi_q}{2\pi} 	\frac{q^6 e^{\pm 4i s \phi_q} e^{i q r \cos(\phi_q - \phi_r)}}{(q^2-(i\omega)^2+\gamma_{01}i\omega)^2(q^2-(i\omega)^2-\gamma_{01}i\omega)^2}  = \int\limits_0^{+\infty} \frac{dq}{2\pi} \frac{ e^{\pm 4 i s \phi_r} q^7 J_4(q r)}{(q^2-(i\omega)^2+\gamma_{01}i\omega)^2(q^2-(i\omega)^2-\gamma_{01}i\omega)^2} \\
&= \frac{e^{\pm 4 i s \phi_r}}{16\pi \gamma_{01}^3(i\omega)^2} \left[2i\omega \left(\Omega_+^4 K_4(\Omega_+ r)-\Omega_-^4 K_4(\Omega_- r) \right) - \gamma_{01} r \left(\Omega_+^5 K_3(\Omega_+ r)+\Omega_-^5 K_3(\Omega_- r) \right)\right] \\
\nonumber I^\pm_{55} &= \int\limits_0^{+\infty} \frac{qdq}{2\pi} \int\limits_0^{2\pi} \frac{d\phi_q}{2\pi} 	\frac{s q^5 e^{\pm 5i s \phi_q} e^{i q r \cos(\phi_q - \phi_r)}}{(q^2-(i\omega)^2+\gamma_{01}i\omega)^2(q^2-(i\omega)^2-\gamma_{01}i\omega)^2}  = \int\limits_0^{+\infty} \frac{dq}{2\pi} \frac{ i s e^{\pm 5 i s \phi_r} q^6 J_5(q r)}{(q^2-(i\omega)^2+\gamma_{01}i\omega)^2(q^2-(i\omega)^2-\gamma_{01}i\omega)^2} = \\
&= \frac{i s e^{\pm 5 i s \phi_r}}{16\pi \gamma_{01}^3(i\omega)^3} \left[2\left(\Omega_+^5 K_5(\Omega_+ r)-\Omega_-^5 K_5(\Omega_- r) \right) + i\omega \gamma_{01} r \left(\Omega_+^4 K_4(\Omega_+ r)+\Omega_-^4 K_4(\Omega_- r) \right)\right]
\end{align}

\section{$T$-matrix calculation}\label{App:Tmatrix}
With the help of $G_0(i\omega,\bs{r})$ and $G_1(i\omega,\bs{r})$ we can calculate the $T$-matrix, whose expression is given by:
\begin{align}
T(i\omega) = \left[\mathbb{I} - V \cdot \lim\limits_{\bs{r} \to \bs{0}} G(i\omega,\bs{r}) \right]^{-1} V.
\end{align}
Here $G(i\omega,\bs{r}) = G_0(i\omega,\bs{r}) + G_1(i\omega,\bs{r})$, and $V$ takes one of the following forms
\begin{align}
\nonumber V_{A1} = U \bpm
		1 & 0 & 0 & 0 \\
		0 & 0 & 0 & 0 \\
		0 & 0 & 0 & 0 \\
		0 & 0 & 0 & 0 
	\epm, 
	\;
V_{B1} = U\bpm
		0 & 0 & 0 & 0 \\
		0 & 1 & 0 & 0 \\
		0 & 0 & 0 & 0 \\
		0 & 0 & 0 & 0 
	\epm, 
V_{A2} = U\bpm
		0 & 0 & 0 & 0 \\
		0 & 0 & 0 & 0 \\
		0 & 0 & 1 & 0 \\
		0 & 0 & 0 & 0 
	\epm, 
	\;
V_{B2} = U\bpm
		0 & 0 & 0 & 0 \\
		0 & 0 & 0 & 0 \\
		0 & 0 & 0 & 0 \\
		0 & 0 & 0 & 1 
	\epm,
\end{align}
with $U$ denoting the magnitude of the impurity potential. At $\bs{r} = 0$ the value of all the integrals that have angular dependence goes to zero, therefore $G_1(i\omega,\bs{r} = \bs{0}) = 0$. Thus in order to determine the value of the $T$-matrix we need to calculate the following:
\begin{align}
\lim\limits_{\bs{r} \to \bs{0}}  G_0(i\omega,\bs{r})  =  \lim\limits_{\bs{r} \to \bs{0}}
	\bpm
		i\omega((i\omega)^2  \negthickspace - \negthickspace \gamma_{01}^2) I_{00} - i\omega I_{03} & 0 & 0 & 0 \\
		0 & (i\omega)^3 I_{00} - i\omega I_{03} & \gamma_{01} (i\omega)^2 I_{00} & 0 \\
		0 & \gamma_{01} (i\omega)^2 I_{00} & (i\omega)^3 I_{00} - i\omega I_{03} & 0 \\
		0 & 0 & 0 & i\omega((i\omega)^2  \negthickspace- \negthickspace \gamma_{01}^2) I_{00} - i\omega I_{03}
	\epm.
\end{align}
The limit above depends on
$
\lim\limits_{r \to +0} I_{00} $ and $\lim\limits_{r \to +0} I_{03}.
$
For the first limit we get: 
\begin{align}
\lim\limits_{r \to +0} I_{00} = -\frac{1}{4\pi \gamma_{01} i\omega } \lim\limits_{r \to +0}  \left[K_0(\Omega_+ r) - K_0(\Omega_- r)\right] = \frac{1}{4\pi \gamma_{01} i\omega }\left(\Omega_+ - \Omega_- \right)
\end{align}
The second limit is given by:
\begin{align}
\nonumber \lim\limits_{r \to +0} I_{03} &= \frac{1}{4\pi \gamma_{01} i\omega} \lim\limits_{r \to +0}  \left[\Omega_+^2 K_0(\Omega_+ r) - \Omega_-^2 K_0(\Omega_- r)\right] = \frac{1}{4\pi \gamma_{01} i\omega} \lim\limits_{r \to +0}  \left[\Omega_+^2 \left(-\gamma_\mathcal{E} - \ln \frac{\Omega_+ r}{2}\right) - \Omega_-^2 \left(-\gamma_\mathcal{E} - \ln \frac{\Omega_- r}{2}\right)\right]  \\
&=  -\frac{1}{4\pi \gamma_{01} i\omega} \lim\limits_{r \to +0} \left[ \gamma_\mathcal{E} \left(\Omega_+^2 - \Omega_-^2 \right) + \Omega_+^2 \ln \frac{\Omega_+ r}{2} - \Omega_-^2 \ln \frac{\Omega_- r}{2}\right],
\end{align}
where $\gamma_\mathcal{E}$ is the Euler–-Mascheroni constant. The expressions above display a logarithmic divergence at $r=0$, and we need to introduce a small-$r$ cutoff. This divergence is a consequence of using a low-energy theory and losing the natural high-energy (small-distance) cutoff of the tight-binding model. The natural cut-off is $r_c=a$, where $a$ is the lattice constant. Since we set the lattice constant to unity above, and all the distances are measured in those units, we can set $r_c = 1$.

Finally, the resulting forms for the T-matrices are given by
\begin{align}
T_{A1} = f(i\omega)\bpm
		1 & 0 & 0 & 0 \\
		0 & 0 & 0 & 0 \\
		0 & 0 & 0 & 0 \\
		0 & 0 & 0 & 0 
	\epm, 
	\;
T_{B1} =g(i\omega) \bpm
		0 & 0 & 0 & 0 \\
		0 & 1 & 0 & 0 \\
		0 & 0 & 0 & 0 \\
		0 & 0 & 0 & 0 
	\epm,
T_{A2} = g(i\omega) \bpm
		0 & 0 & 0 & 0 \\
		0 & 0 & 0 & 0 \\
		0 & 0 &  1 & 0 \\
		0 & 0 & 0 & 0 
	\epm, 
	\;
T_{B2} = f(i\omega) \bpm
		0 & 0 & 0 & 0 \\
		0 & 0 & 0 & 0 \\
		0 & 0 & 0 & 0 \\
		0 & 0 & 0 & 1 
	\epm,
\end{align}
where we defined
\begin{align}
f(i\omega) \equiv \frac{U}{1 - U \lim\limits_{r \to +0} \left[ i\omega((i\omega)^2-\gamma_{01}^2) I_{00} - i\omega I_{03} \right] },\;
 g(i\omega) \equiv \frac{U}{1 - U \lim\limits_{r \to +0} \left[ (i\omega)^3 I_{00} - i\omega I_{03} \right] }.
\end{align}

\section{Local Density of States}\label{App:LDOS}

\subsection{Real space}

Asymptotically at large $r$ we have:
\begin{align}
\nonumber \delta\rho_{A1}(\epsilon,r,\phi_r) &= -\frac{1}{\pi} \im \left\{ \frac{f(i\omega)}{32 \pi\, i\omega}  \left[ \gamma_{01} \left(\Omega_+ e^{-2 \Omega_+ r}-\Omega_- e^{-2 \Omega_- r}\right) - 2i\omega \left( \gamma_{01}^2 - i\omega \left( i\omega - i \sqrt{\gamma_{01}^2 - (i\omega)^2} \right) \right) \frac{e^{-(\Omega_++\Omega_-)r }}{\sqrt{\Omega_-} \sqrt{\Omega_+}} \right]  \frac{1}{r} \right. \\
\label{App:eq:rhoA1}&\left. \phantom{aaaaaaaa} + \gamma_{03} \frac{f(i\omega)}{32 \pi\, i\omega} \left[ \Omega_+^2 e^{-2 \Omega_+ r}-\Omega_-^2 e^{-2 \Omega_- r} + \sqrt{\Omega_+} \sqrt{\Omega_-} (\Omega_+ - \Omega_-) e^{-(\Omega_+ + \Omega_-) r} \right] \frac{\sin 3\phi_r}{r} \right\}, \\
\label{App:eq:rhoB1} \delta\rho_{B1}(\epsilon,r,\phi_r) &= -\frac{1}{\pi} \im \left\{ \frac{ g(i\omega)}{32 \pi}  \left[ i\omega \gamma_{01} \left(\frac{e^{-2 \Omega_+ r}}{\Omega_+} - \frac{e^{-2 \Omega_- r}}{\Omega_-}\right) + 2 \left((i\omega)^2 + \Omega_+ \Omega_- \right) \frac{ e^{-(\Omega_++\Omega_-)r }}{\sqrt{\Omega_+} \sqrt{\Omega_-}} \right]  \frac{1}{r} \right. \\
\nonumber &\left.  - \gamma_{03} \frac{g(i\omega)}{32 \pi} \left[ (\gamma_{01} - i\omega) e^{-2 \Omega_+ r} + (\gamma_{01} + i\omega) e^{-2 \Omega_- r} + \left( (\gamma_{01} + i\omega)\Omega_+ + (\gamma_{01} - i\omega) \Omega_- \right) \frac{ e^{-(\Omega_+ + \Omega_-) r}}{\sqrt{\Omega_+} \sqrt{\Omega_-}} \right] \frac{\sin 3\phi_r}{r} \right\}, \\
\nonumber \delta\rho_{A2}(\epsilon,r,\phi_r) &= -\frac{1}{\pi} \im \left\{ \frac{ g(i\omega)}{32 \pi}  \left[ i\omega \gamma_{01} \left(\frac{e^{-2 \Omega_+ r}}{\Omega_+} - \frac{e^{-2 \Omega_- r}}{\Omega_-}\right) - 2 \left((i\omega)^2 + \Omega_+ \Omega_- \right) \frac{ e^{-(\Omega_++\Omega_-)r }}{\sqrt{\Omega_+} \sqrt{\Omega_-}} \right]  \frac{1}{r} \right. \\
\nonumber &+ \gamma_{03} \frac{g(i\omega)}{32 \pi \gamma_{01}^2} 
\Bigg[ \left(\gamma_{01}^3 - 2i \omega \gamma_{01}^2 + 2(i\omega)^2 \gamma_{01} - (i\omega)^3 \right) e^{-2\Omega_+ r} + \left(\gamma_{01}^3 + 2i \omega \gamma_{01}^2 + 2(i\omega)^2 \gamma_{01} + (i\omega)^3 \right) e^{-2\Omega_- r} \\
\nonumber &\left. -\Big(\left(\gamma_{01}^3 + 4i \omega \gamma_{01}^2 + 2(i\omega)^2 \gamma_{01} - (i\omega)^3 \right) \Omega_+ + \left(\gamma_{01}^3 - 4i \omega \gamma_{01}^2 + 2(i\omega)^2 \gamma_{01} + (i\omega)^3 \right) \Omega_- \Big) \frac{e^{-(\Omega_+ + \Omega_-) r}}{\sqrt{\Omega_+}\sqrt{\Omega_-}} \right] \frac{\sin 3\phi_r}{r} \\
\label{App:eq:rhoA2}&\left. +\gamma_{03} \frac{g(i\omega)}{8 \pi \gamma_{01}^3} i\omega \left(\gamma_{01}^2-(i\omega)^2 \right)\left[\frac{\gamma_{01} - i\omega}{\Omega_+} e^{-2\Omega_+ r} - \frac{\gamma_{01} + i\omega}{\Omega_-} e^{-2\Omega_- r} + 2 i\omega \frac{e^{-(\Omega_+ + \Omega_-)r}}{\sqrt{\Omega_+}\sqrt{\Omega_-}} \right] \frac{\sin 3\phi_r}{r^2}\right\}, \\
\nonumber \delta\rho_{B2}(\epsilon,r,\phi_r) &= -\frac{1}{\pi} \im \left\{ \frac{f(i\omega)}{32 \pi}  \Bigg[ \gamma_{01}\left(\frac{\gamma_{01} - i\omega}{\Omega_+} e^{-2\Omega_+ r} + \frac{\gamma_{01} + i\omega}{\Omega_-} e^{-2\Omega_- r} \right) +2 \left(\gamma_{01}^2 -(i\omega)^2 - \Omega_+ \Omega_- \right) \frac{e^{-(\Omega_++\Omega_-)r }}{\sqrt{\Omega_+}\sqrt{\Omega_-} }  \Bigg] \frac{1}{r} \right. \\
\label{App:eq:rhoB2}&\left. -\gamma_{03} \frac{f(i\omega)}{32\pi} \Bigg[ (\gamma_{01} - i\omega) e^{-2 \Omega_+ r} + (\gamma_{01} + i\omega) e^{-2 \Omega_- r} + 3 \left( (\gamma_{01} + i\omega) \Omega_+ + (\gamma_{01} - i\omega) \Omega_- \right)  \frac{e^{-(\Omega_++\Omega_-)r }}{\sqrt{\Omega_+}\sqrt{\Omega_-}} \Bigg] \frac{\sin 3\phi_r}{r}  \right\}.
\end{align}

\subsection{Momentum space}
Below we assume that the energy $\epsilon > 0$, and hence the terms with $\Omega_-$ in Eqs.~(\ref{eq:rhoA1}-\ref{eq:rhoB2}) provide the dominant contribution into the asymptotic behavior of the local density of states. If we chose negative energies, i.e., $\epsilon < 0$, then the terms with $\Omega_+$ would dominate. To perform Fourier transforms, we will use the two following integrals:
\begin{align}
\nonumber \mathcal{F}_0(p,\Omega) &\equiv \mathcal{F}\left[\frac{e^{-\Omega r}}{r} \right] = \int d\bs{r} \frac{e^{-\Omega r}}{r} e^{-i \bs{p}\cdot \bs{r}} = \int\limits_0^\infty dr e^{-\Omega r} \int\limits_0^{2\pi} d\phi_r e^{-i p r \cos(\phi_r - \phi_p)} = \\
&= 2\pi \int\limits_0^\infty dr\, J_0(pr) e^{-\Omega r} = \frac{2\pi}{\Omega} \frac{1}{\sqrt{1+p^2/\Omega^2}} \\
\nonumber \mathcal{F}_1(p,\phi_p,\Omega) &\equiv \mathcal{F}\left[\frac{e^{-\Omega r}}{r} \sin 3\phi_r \right] = \int d\bs{r} \frac{e^{-\Omega r}}{r} \sin 3\phi_r\, e^{-i \bs{p}\cdot \bs{r}} = \int\limits_0^\infty dr e^{-\Omega r} \int\limits_0^{2\pi} d\phi_r\, \sin 3\phi_r\, e^{-i p r \cos(\phi_r - \phi_p)} = \\
&= 2\pi i \sin 3\phi_p \int\limits_0^\infty dr\, J_3(pr) e^{-\Omega r} = 2\pi i \sin 3\phi_p \frac{p^2\left(-3+ \sqrt{1+p^2/\Omega^2} \right) + 4\Omega^2\left(-1+ \sqrt{1+p^2/\Omega^2} \right)}{p^3\sqrt{1+p^2/\Omega^2}},
\end{align}
where $(p, \phi_p)$ are the polar coordinates in momentum space. Thus the Fourier transforms read
\begin{align}
\label{App:eq:rhoA1FT} \delta\rho_{A1} &= \frac{1}{2\pi i} \frac{ f(\epsilon) \Omega_- \left[ \gamma_{01} \mathcal{F}_0(p,2\Omega_-) + \gamma_{03} \Omega_- \mathcal{F}_1 (\bs{p}, 2\Omega_-) \right] -  f^*(\epsilon) \Omega^*_- \left[ \gamma_{01} \mathcal{F}_0(p,2\Omega^*_-) + \gamma_{03} \Omega^*_- \mathcal{F}_1(\bs{p}, 2\Omega^*_-) \right]}{32 \pi \epsilon}, \\
\label{App:eq:rhoB1FT} \delta\rho_{B1} &= \frac{1}{2\pi i} \gamma_{01} \frac{ \frac{g(\epsilon)}{ \Omega_-} \left[ \epsilon \mathcal{F}_0(p,2\Omega_-) + \gamma_{03} \Omega_- \mathcal{F}_1 (\bs{p}, 2\Omega_-) \right] -  \frac{g^*(\epsilon)}{ \Omega^*_-} \left[ \epsilon \mathcal{F}_0(p,2\Omega^*_-) + \gamma_{03} \Omega^*_- \mathcal{F}_1(\bs{p}, 2\Omega^*_-) \right]}{32 \pi}, \\
\label{App:eq:rhoA2FT} \delta\rho_{A2} &= \frac{1}{2\pi i} \gamma_{01} \frac{ \frac{g(\epsilon)}{ \Omega_-} \left[ \epsilon \mathcal{F}_0(p,2\Omega_-) - \gamma_{03} \Omega_- \mathcal{F}_1 (\bs{p}, 2\Omega_-) \right] -  \frac{g^*(\epsilon)}{ \Omega^*_-} \left[ \epsilon \mathcal{F}_0(p,2\Omega^*_-) - \gamma_{03} \Omega^*_- \mathcal{F}_1(\bs{p}, 2\Omega^*_-) \right]}{32 \pi}, \\
\label{App:eq:rhoB2FT}  \delta\rho_{B2} &= - \frac{1}{2\pi i} \gamma_{01} \frac{ \frac{f(\epsilon)}{ \Omega_-} \left[ \gamma_{01} \mathcal{F}_0(p,2\Omega_-) - \gamma_{03} \Omega_- \mathcal{F}_1 (\bs{p}, 2\Omega_-) \right] -  \frac{f^*(\epsilon)}{ \Omega^*_-} \left[ \gamma_{01} \mathcal{F}_0(p,2\Omega^*_-) - \gamma_{03} \Omega^*_- \mathcal{F}_1(\bs{p}, 2\Omega^*_-) \right]}{32 \pi}.
\end{align}
From the equations above we find the positions of ring-like resonances in momentum space. Indeed, both functions $\mathcal{F}_0(p,2\Omega_-)$ and $\mathcal{F}_1(p,\phi_p,2\Omega_-)$ have resonances at
\begin{align}
p_{\mathrm{res}} = - 2 i \Omega_- = 2 \sqrt{\epsilon(\gamma_{01}+\epsilon)}.
\end{align}
If we chose negative energies, i.e., $\epsilon < 0$, then the resonances would appear at $p_{\mathrm{res}} = - 2 i \Omega_+ = 2 \sqrt{|\epsilon|(\gamma_{01}+|\epsilon|)}$.

\end{document}